\documentclass[a4paper,fleqn]{cas-dc}

\usepackage[toc,page]{appendix}
\usepackage{amsmath}
\usepackage{algorithm2e}
\usepackage{kantlipsum}
\usepackage{multirow}
\usepackage{mathtools}
\usepackage{palatino}
\usepackage{verbatim} 
\usepackage{amsmath,amssymb,amsthm}
\usepackage{mathtools}
\usepackage{physics}
\usepackage{algpseudocode}
\usepackage{enumitem,kantlipsum}

\usepackage{caption}
\usepackage{graphicx}
\usepackage[numbers]{natbib}
\usepackage[utf8x]{inputenc}
\usepackage[T2A]{fontenc}
\usepackage{lipsum}
\usepackage[english]{babel}
\usepackage{kantlipsum}
\usepackage[nodisplayskipstretch]{setspace} 

\usepackage{wrapfig}
\usepackage{lscape}
\usepackage{rotating}
\usepackage{epstopdf}
\usepackage[ragged]{sidecap}  
\usepackage[english]{babel}

\AtBeginDocument{%
  \addtolength\abovedisplayskip{-0.5\baselineskip}%
  \addtolength\belowdisplayskip{-0.5\baselineskip}%
}

\usepackage{longtable}
\usepackage{threeparttable}
\usepackage{bigstrut}
\usepackage{multirow}
\usepackage{multirow}
\usepackage{rotating}
\usepackage{pifont}
\usepackage[
singlelinecheck=false 
]{caption}

\usepackage{fancyhdr}
\usepackage{etoolbox}
\makeatletter
\patchcmd{\ps@pprintTitle}{\footnotesize\itshape
       XXXPreprint submitted to \ifx\@journal\@empty Elsevier
       \else\@journal\fi\hfill\today}{\relax}{}{}
\makeatother

\makeatletter
\def\ps@pprintTitle{%
 \let\@oddhead\@empty
 \let\@evenhead\@empty
 \def\@oddfoot{\centerline{\thepage}}%
 \let\@evenfoot\@oddfoot}
\makeatother

\def\tsc#1{\csdef{#1}{\textsc{\lowercase{#1}}\xspace}}
\tsc{WGM}
\tsc{QE}
\tsc{EP}
\tsc{PMS}
\tsc{BEC}
\tsc{DE}

\begin{document}
\pagestyle{fancy}
\fancyhf{} 
\rhead{} 
\lhead{}
\rfoot{Page \thepage}

\shorttitle{HARDC : A novel ECG-based heartbeat classification method to detect arrhythmia using hierarchical attention based dual structured RNN with dilated CNN} 
\shortauthors{Accepted in Neural Networks, elsevier (March 2023), Copyright belongs to the Journal, Shofiqul et~al.}

\title [mode = title]{HARDC : A novel ECG-based heartbeat classification method to detect arrhythmia using hierarchical attention based dual structured RNN with dilated CNN}             \author[1,2]{Md Shofiqul Islam}

\address[1]{Faculty of Computing, Universiti Malaysia Pahang, Gambang-26300, Kuantan, Pahang, Malaysia.}
\address[2]{IBM Centre of Excellence , Centre for Software Development \& Integrated Computing, Universiti Malaysia Pahang (UMP), Lebuhraya Tun Razak,Gambang-26300, Kuantan, Pahang, Malaysia.}

\author[3]{Khondokar Fida Hasan}
\address[3]{School of Computer Science, Queensland University of Technology (QUT), 2 George Street, Brisbane 4000, Australia}

\author[4]{Sunjida Sultana}

\address[4]{Department of Computer Science and Engineering, Islamic University, Kushtia-7600, Bangladesh}

\author[5]{Shahadat Uddin}
\address[5]{School of Project Management, Faculty of Engineering, The University of Sydney, Sydney, Australia.}

\author[6]{Pietro Lio'}
\address[6]{Department of Computer Science and Technology, University of Cambridge, Cambridge, UK.}


\author[7]{Julian M.W. Quinn}
\address[7]{Bone Research Group, The Garvan Institute of Medical Research,  Darlinghurst, NSW Australia}

\author[8]
{Mohammad Ali Moni}[type=editor,
                        auid=004,bioid=4,
                       orcid=0000-0003-0756-1006
                       ]
\cormark[1]
\ead{m.moni@uq.edu.au}

\address[8]{Artificial Intelligence \& Data Science, School of Health and Rehabilitation Sciences, Faculty of Health and Behavioural Sciences, The University of Queensland St Lucia, QLD 4072, Australia.}

\nonumnote{Corresponding author:
  }

\cortext[cor]{Mohammad Ali Moni}


\begin{abstract}
Deep learning-based models have achieved significant success in detecting cardiac arrhythmia by analyzing ECG signals to categorize patient heartbeats. 
To improve the performance of such models, we have developed a novel hybrid hierarchical attention-based bidirectional recurrent neural network with dilated CNN (HARDC) method for arrhythmia classification.
This solves problems that arise when traditional dilated convolutional neural network (CNN) models disregard the correlation between contexts and gradient dispersion. The proposed HARDC fully exploits the dilated CNN and bidirectional recurrent neural network unit (BiGRU-BiLSTM) architecture to generate fusion features. As a result of incorporating both local and global feature information and an attention mechanism, the model's performance for prediction is improved.
By combining the fusion features with a dilated CNN and a hierarchical attention mechanism, the trained HARDC model showed significantly improved classification results and interpretability of feature extraction on the PhysioNet 2017 challenge dataset. Sequential Z-Score normalization, filtering, denoising, and segmentation are used to prepare the raw data for analysis. CGAN (Conditional Generative Adversarial Network) is then used to generate synthetic signals from the processed data.
The experimental results demonstrate that the proposed HARDC model significantly outperforms other existing models, achieving an accuracy of 99.60\%, F1 score of 98.21\%, a precision of 97.66\%, and recall of 99.60\% using MIT-BIH generated ECG. In addition, this approach substantially reduces run time when using dilated CNN compared to normal convolution. Overall, this hybrid model demonstrates an innovative and cost-effective strategy for ECG signal compression and high-performance ECG recognition. Our results indicate that an automated and highly computed method to classify multiple types of arrhythmia signals holds considerable promise. 


\end{abstract}


\maketitle
\section{Introduction}
 Electrocardiogram (ECG) records are a commonly used diagnostic method used in monitoring heart activity. The information in ECG recordings can indicate heart irregularities and facilitate diagnoses of cardiac pathologies, such as arrhythmia, ischemic heart conditions, and the occurrence of myocardial infarction \cite{1yildirim2018arrhythmia}. Arrhythmia involves irregularities in  heartbeat rate due to electrical conduction issues, while atrial fibrillation (AF) is a persistent and particularly dangerous type of arrhythmia with very rapid cardiac contractions. AF and other types of arrhythmias can be monitored using ECG signals \cite{2oh2018automated}. To identify arrhythmias and their severity, clinical assessment of patients and their ECG signals by health professionals is the current practice. However, automated analysis solutions are beginning to emerge. \cite{3cano2017essential}. 
 
A machine learning-based approach to automated detection of ECG signals requires feature detection and extraction, followed by classification. However, many automated categorization techniques face challenges in achieving detection with sufficient accuracy\cite{4dang2019novel}  \cite{5chen2020multi}. This also includes issues with ECG signal classification  and labeling \cite{6jiang2019novel}, data annotation \cite{7hong2020opportunities} , interpretability \cite{zhang2021interpretable} \cite{chou2020knowledge} \cite{8li2019interpretability}, efficiency \cite{9hinton2015distilling}, labeling of class correctly, and handling multimodal data \cite{7hong2020opportunities} \cite{10hammad2018multimodal}. Other specific challenges that arise during ECG signal preprocessing are standardization, handling of various ECG signal patterns, waveform variability in ECG signals, and long-length ECG signal processing.

ECG signal identification currently requires expert opinion to detect and diagnose arrhythmia, but various hybrid approaches to this detection problem have been proposed, which involve machine learning methods \cite{11wang2021intelligent} \cite{4dang2019novel} \cite{5chen2020multi}. However, deep learning-based methods have generally outperformed classic machine learning techniques \cite{18schuster1997bidirectional} \cite{5chen2020multi} \cite{2oh2018automated} \cite{13yildirim2018efficient} \cite{humayn2021explainable} \cite{hasan2022security} and produced excellent results with respect to precision and performance. Deep learning methods use algorithm stages in a model construction trained with the underlying data properties. Deep learning techniques, especially convolutional neural networks (CNN) and recurrent neural networks (RNN), which primarily contain long short-term memory (LSTM) and bidirectional LSTM (Bi-LSTM) networks, have been particularly successful in ECG signal processing \cite{11wang2021intelligent}. 

CNN is a common technique in deep learning studies \cite{1yildirim2018arrhythmia} \cite{22chen2017heartbeat} \cite{21ye2012heartbeat} \cite{23mar2011optimization} \cite{24shi2019hierarchical} \cite{mim2023gru} \cite{islam2022cnn} \cite{bala2022efficient} because it performs both feature and classification processes with a traditional training stage. Several convolutional and pooling layers are cascaded in the CNN model to form a deep network capable of extracting inherent input features. Because CNN structures have proven to be effective in computationally demanding studies, they are a good fit for ECG signal recognition mechanisms \cite{22chen2017heartbeat}\cite{21ye2012heartbeat}\cite{23mar2011optimization}. Acharya et al. \cite{27elhaj2016arrhythmia} constructed a CNN architecture with 11 layers that automatically identify signatures that reveal shockable and non-shockable ventricular arrhythmias in segmented ECG signals. CNN has also been utilized for automatic arrhythmia detection at varying intervals of heartbeat ECG segments \cite{20waldo2008inter}. Yildirim et al. \cite{1yildirim2018arrhythmia} employed a 16-layer deep CNN model to categorize extended-duration ECG data. Another important method that has been widely used in recent deep learning-based work is the LSTM \cite{2oh2018automated} networks. The categorization of ECG signals has also been performed with LSTMs \cite{5chen2020multi} and several approaches for enhancing the efficiency of LSTM networks have been developed. LSTM and CNN have also been integrated into recent work to provide a hybrid method for arrhythmia identification using variable-length heartbeats \cite{2oh2018automated}. Tan et al. \cite{21ye2012heartbeat} used an LSTM network employing CNN that was able to detect coronary artery disease from ECG signal recordings automatically and with acceptable accuracy. Using only wavelet sequencing as the network input can significantly boost the performance of LSTM-based algorithms \cite{1yildirim2018arrhythmia}.

RNN experiences vanishing gradient problems and difficulties in handling long sequence dependency among features \cite{16andersen2019deep}\cite{17hochreiter1997long}. However, as previous LSTM networks have only been able to model such sequences in one way, Bi-LSTM was developed to analyze the pattern in both forward as well as backward directions simultaneously. Although Bi-LSTM generally outperforms LSTM in various settings, multiple investigations have revealed that it suffers from problems associated with data duplication. Schuster et al. \cite{18schuster1997bidirectional} showed that while Bi-LSTM performs well for some applications, this work did not demonstrate how to use and combine feature representations with the forward and backward options efficiently and has a problem of information redundancy in the network. According to Pascanu et al. \cite{19schuster1997bidirectional}, Bi-LSTM is prone to information redundancy with gradient explosion problems, making it unsuitable for lightweight devices. 

\begin{figure*}[t]
\centering
\includegraphics[width=.71\textwidth]{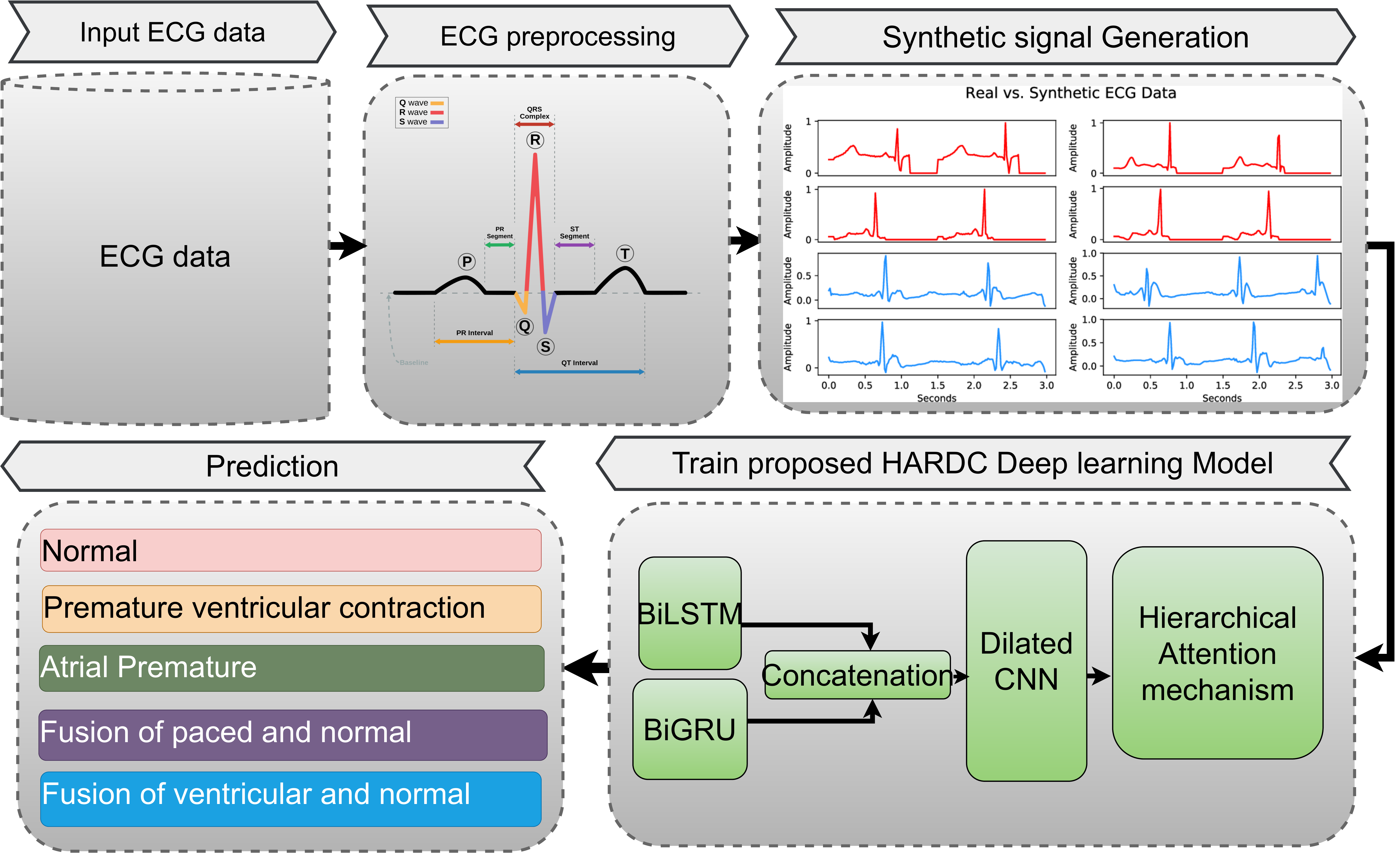}
\caption{Overall basic flowchart of HARDC model.}
\label{fig:flochar overall}
\end{figure*}

We employed an arrhythmia-diagnosis method employing hybrid deep learning to detect AF automatically from the ECG dataset. Our proposed method consists of data processing, synthetic signal generation, feature extraction, and classification. It combines four primary stages: Synthetic ECG generation with a conditional generative adversarial network (CGAN) done after the preprocessing; Bidirectional RNN; Dilated CNN; and hierarchical attention layer to extract features and classify ECG.  The first stage of preprocessing requires peak detection and data segmentation to collect the input data. This is then input to the second stage, which employs dilated convolution layers and two Bidirectional Long-short Term memory(BiLSTM)-Bidirectional Gated Recurrent Unit(BiGRU) (BiLSTM-BiGRU) blocks to accomplish automatic feature extraction. The ECG signal episodes are then input to a dilated convolutional network with two completely connected layers with a SoftMax layer, which has two fully connected layers. The generated output is then forwarded to the hierarchical attention layer for final categorization after the operation of the Dilated CNN layer.
The basic steps of our proposed methodology are presented using the flowchart in figure \ref{fig:flochar overall}.

Finally, we have performed tests to evaluate the performance of the proposed method. Utilizing the extensive-ranging extracted features from the BiLSTM-BiGRU as well as the speedily extracted relevant feature from the dialed CNN layer, this hierarchical attention method improves feature extraction interpretability. This improves the prediction accuracy of ECG signal categorization.

In summary, the primary contributions of this paper are, first, demonstrating an improved BiLSTM-BiGRU method enabling the capture of variable-sized and long-ranged features important for ECG signal analysis. Second, the development of an updated hierarchical attention method to optimize feature weights that improve ECG feature selection interpretability and enhance ECG signal classification results. Finally, we propose a novel dilated CNN to handle more complex features with fewer network parameters that are able to increase the classification performance with less complexity. Our method eliminates the requirement for traditional manual feature extraction methods while delivering accurate results with improved interpretability, so it constitutes a novel method for classifying ECG signals.

\section{	Related Study  }
There is an increased risk of cardiovascular disease associated with an irregular cardiac rhythm. In addition, atrial fibrillation, obesity, hypertension, atrial-premature or tricuspid valve diseases, obstructive sleep apnea, hyperthyroidism, and alcoholism similarly reduce the quality of life  and raise the risk of mortality from cardiovascular disease. Atrial fibrillation is a known arrhythmia disease that is  necessary to diagnose, manage, and avoid related consequences \cite{4dang2019novel}.
Some existing traditional machine learning methods \cite{13yildirim2018efficient}\cite{14acharya2017automated} have been shown to perform well for ECG classification in recent times. SVM classifiers are a commonly used method among machine learning methods. SVM classifiers with Wavelet coefficient and RR interval(in ECG, the time between two consecutive R-waves of a QRS signal)  for 0.83s long ECG signal classification for a variable number of samples have been proposed by Ye and Coimbra \cite{21ye2012heartbeat}. Another SVM classifier with RR and ECG projection matrix to classify variable-length ECG signals was developed by Chen et al. \cite{22chen2017heartbeat}. LDA-MLP(Linear Discriminant Analysis (LDA)-Multilayer Perception) \cite{23mar2011optimization} method with temporal and morphological features, XGBoost with handcraft features \cite{24shi2019hierarchical}, has also been used for ECG classification. RR intervals, wavelet coefficients,  and morphological descriptors have also been employed in SVM-based ECG classification \cite{25mondejar2019heartbeat}. LS SVM  classifiers with PCA(Principal Component Analysis), DWT(Discrete Wavelet Transform), HOS(Higher Order Spectra), and ICA features have been developed to classify ECG signals in the work of Martis et al.This work used more complex features, and it requires higher computational cost.  \cite{26martis2012application}. 
This section delves into the methodologies directly related to this research work described here, i.e., the machine learning and deep learning approaches applied to detect AF using ECG signals. Note that space does not permit us to describe all of the methods we have employed here.

\subsection{	Traditional Machine Learning}
There are some existing traditional machine learning methods. SVM \cite{21ye2012heartbeat} \cite{22chen2017heartbeat} \cite{26martis2012application} \cite{23mar2011optimization} \cite{25mondejar2019heartbeat}\cite{27elhaj2016arrhythmia} \cite{28yang2018automatic}, XGBoost \cite{24shi2019hierarchical}, Random Forest \cite{29li2016ecg} and LDA-MLP \cite{23mar2011optimization} are the ones showing improved performance in ECG classification in recent times. These methods take more training time, and the obtained accuracy is lower than other deep learning methods. SVM classifier is a  commonly used method among machine learning approaches. A method using SVM classifier that uses Wavelet coefficient as well as RR interval to classify 0.83s long ECG signal \cite{21ye2012heartbeat}. This method got 89.02\% accuracy, 62.2\% precision, and 56.27\% recall, but this method is very dependent on feature engineering. Another SVM algorithm using RR and the projection matrix was  classification of variable-length ECG signal \cite{22chen2017heartbeat}. This method obtained 98.46\% accuracy in classifying fifteen different classes of 0.83s long ECG, but this method has a bit higher time and space complexity. Another method used XGBoost algorithms that work with handcraft features \cite{24shi2019hierarchical} to the classify ECG, but it does not perform well in the prediction of abnormal class. A simple SVM method that used RR intervals, Wavelet coefficient, and Morphological to detect arrhythmia \cite{25mondejar2019heartbeat}. Most of the SVM classifiers \cite{2oh2018automated}\cite{25mondejar2019heartbeat}\cite{21ye2012heartbeat}  to classify ECG signals got accuracy between 91\% to 97\% but most of them are very dependent on feature engineering. SVM \cite{27elhaj2016arrhythmia} algorithm that uses PCA, DWT,HOS, ICA features in classification of ECG signal to obtain the highest accuracy 98.91\%. This method is comparatively complex in computation for its higher processing of features.
\subsection{	Deep Learning   }
Traditional methods of machine learning can be faster but can not handle big datasets \cite{2oh2018automated}\cite{25mondejar2019heartbeat}\cite{21ye2012heartbeat}. The most common deep learning-based method is CNN. It showed its performance with more excellent accuracy in classification \cite{31zubair2016automated}\cite{30kiranyaz2015real} than other machine \cite{2oh2018automated}\cite{25mondejar2019heartbeat}\cite{21ye2012heartbeat} and deep learning methods \cite{33yildirim2019new}\cite{4dang2019novel}. We listed some basic deep learning techniques to detect arrhythmia from MIT-BIH (Massachusetts Institute of Technology-Beth Israel Hospital) database. Another CNN method  \cite{30kiranyaz2015real} that used patient-specific features to detect arrhythmia. This technique predicts arrhythmia with 99\% precision. The main limitation of this method is the characterization of the crucial anomaly beats. Another CNN method \cite{31zubair2016automated} got 92.70\% accuracy in classifying 0.5s long ECG signals with 200 samples. Bagging algorithms and biased loss filtering are the main limitations of this method. Yildirim et al. \cite{13yildirim2018efficient} developed a comparatively new and popular CNN-based method with synthetic data generation and rescaling raw data to classify ECG signal with 260 samples. This method achieved an accuracy of 94.03\% while classifying 360 samples with a duration of 1s. Another CNN method used random projections with RR interval \cite{32sellami2019robust} to classify heartbeat. This method gave 88.33\% accuracy, 64.20\% precision, and 57\% recall to classify variable-length heartbeat signals.
\subsection{	Hybrid Deep Learning   }
Hybrid deep learning methods emerged with improved performance and adaptability over conventional methods. The hybrid deep learning-based method is the combination of different general deep learning techniques.
Another hybrid method utilized CNN with RNN and used patient-based features to detect arrhythmia. This method got 98.10\% accuracy on variable-length ECG data \cite{2oh2018automated}. This method achieves slightly low accuracy than normal SVM \cite{27elhaj2016arrhythmia} . Yıldırım et al. \cite{1yildirim2018arrhythmia} developed a method using Deep CNN for arrhythmia detection with rescaling of signals. This method classifies small length (0.015s) ECG obtaining 91.33\% precision. Additionally, the same authors developed a method using attention-based LSTM for handling code-based signal \cite{33yildirim2019new}, which gave an accuracy of 99.11\%  but worked with short-length ECG data. Another hybrid model utilizes CNN with BiLSTM and uses four RR interval information \cite{5chen2020multi} to classify ECG signals. Accuracy of 96.77\%, 77.8\% precision, and 81.2\% recall was obtained by this method. Its performance should be more adaptable to different data.  Dilated CNN is developed by Ma et al. \cite{12ma2021ecg}. In this method, denoised and normalized data are used to detect arrhythmia. This technique achieved 98.65\% precision. This technique reduces running time, but it requires more normalization with some prepossessing work that sometimes leads to the wrong predictions. Dang et al. \cite{4dang2019novel} developed a deep CNN with BiLSTM to classify ECG signal analyzing features of RR intervals, PR intervals, and QRS (QRS represents ventricular depolarization with the integration of 'Q' wave, 'R' wave, and 'S' wave). It categorizes 100 samples of ECG signals into five predictions of class. This approach achieved 96.59\% in classifying five types of ECG data. The main limitation is that it requires more time in execution and has no generation ability in real-time usability.  Jin et al. \cite{34jin2020multi} developed a technique that used attention with CNN-LSTM in analyzing multi-domain features. It obtained 98.51\% accuracy to predict five different ECG signals, but its adaptability is weak. A new technique using CNN-BiLSTM \cite{11wang2021intelligent} focuses on raw data features of 360 samples. This method obtained a very high 99.11\% accuracy, however, its generalisability could be improved.

A multivariate ECG classification by residual channel attention networks \cite{C1tao2022ecg} and BiGRU got 96.7\% and 97.7\% accuracy for long and short-length ECG signals, respectively. This method used lead attention, morphological and temporal features nicely. Limitation of this method is that it does not show the real-time data analysis result. Its gained accuracy is also slightly lower than some related methods\cite{2oh2018automated} \cite{30kiranyaz2015real}. 

\subsection{Research Motivation}

Existing methods and their limitations motivated us to develop a new and more accurate ECG classification approach.
Traditional machine learning \cite{21ye2012heartbeat} \cite{22chen2017heartbeat} \cite{26martis2012application} \cite{27elhaj2016arrhythmia} \cite{28yang2018automatic} \cite{23mar2011optimization} algorithms gain almost 1\% down accuracy from 99\% than conventional deep learning(CNN) method\cite{2oh2018automated} \cite{25mondejar2019heartbeat} \cite{21ye2012heartbeat} when use temporal features. Performance in achieving accuracy remains almost similar when wavelet features are used in (Support Vector Machine (SVM)-based methods\cite{27elhaj2016arrhythmia}. 
Synthetic signal generation can also balance the signal sufficiently \cite{2oh2018automated}]. 
Based on a recent review, we see that the attention-based model\cite{C1tao2022ecg} can capture the most relevant features, and dilated CNN \cite{5chen2020multi} \cite{33yildirim2019new} \cite{4dang2019novel}\cite{4dang2019novel} can extract features at a faster speed \cite{12ma2021ecg}. 
This further motivated us to develop a novel method for ECG classification using attention dilated CNN with Bidirectional RNN. Our reprocessed tasks of generating a synthetic signal with a wavelet and QRS features slightly boost the performance as well.

We concluded that the existing general machine learning(ML) methods of ECG classification are quicker but not as accurate and reproducible as deep learning and hybrid deep learning. However, all methods show some problems, having issues such as levels of feature extraction interpretability,  running time, higher memory requirement, lack of adaptability or generalization, lower accuracy, and high loss in prediction. We addressed these issues with the proposed method developed as shown below.

\section{Data Information}
For detecting arrhythmic heartbeats, a new attention-based BiGRU-BiLSTM with Dilated CNN technique was developed in this study. The HARDC approach was utilized to compress and synthesize ECG beats inside the proposed system. Each ECG trace was first converted into low-dimensional encoded signals. A BiGRU - BiLSTM network model extracts features, and then the RNN layer's output is received by a Dilated CNN. Hierarchical attention was employed to classify the synthetic signals in the final stage. Following that, arrhythmic beats and synthetic signal recognition were done. The experiments used 109446 segmented signals from five different ECG beat types.
\subsection{Dataset}
\begin{figure*}[t]
\centering
\includegraphics[width=0.9\textwidth]{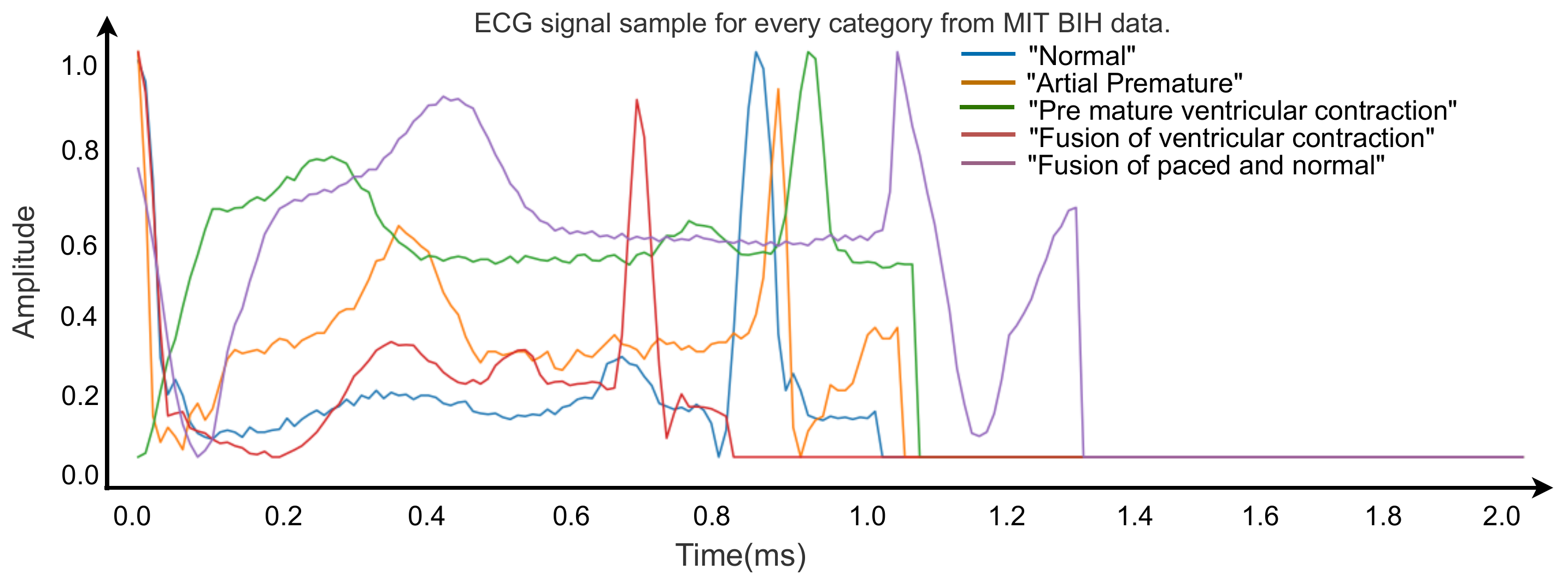}
\caption{Arrhythmia types with the sample.}
\label{fig:ecg type}
\end{figure*}

Our implemented data is taken from the PhyoNet website and it is publicly available. Cardiologist annotated the ECG beat, and the diagnoses were settled by consensus. ECG recordings from the MIT-BIH database \cite{35moody2001physionet} provide the basis of our approach. The MIT-BIH arrhythmia data is used in predicting arrhythmia. For ECG classification techniques, this is regarded as the gold-standard evaluation data \cite{35moody2001physionet}. The database has 48 half-hour ECG records from patients investigated at the BIH Arrhythmia Laboratory. Each record includes two 30-minute ECG lead signals at 360 samples per second(s). Manually, data annotation and checking were done for both the heartbeat classification method with timing information. A total of 109,492 heartbeats were tagged. We looked at five types of ECG signals in this study to detect arrhythmia. Table \ref{table data amount} displays the different types of arrhythmias and how much amount per class. Table 1 presents the amount of signal per class. This table clearly shows the "Normal" class with a higher amount of 82.8\% and the lower amount for the ECG class "Fusion of paced and Normal". 
\begin{table}
\caption{Arrhythmia types}
\label{table data amount}
\small
\begin{tabular}{ |p{5cm}|p{1cm}|p{1cm}|}
\hline
\multicolumn{3}{|c|}{Data Information} \\
\hline
ECG class & Sample Size & Percent\\\hline
 'Normal'  & 90589 & 82.8\%\\\hline
 'Fusion of paced and normal'  & 8039 & 07.3\%\\\hline
 'Premature ventricular contraction'  & 7236 & 06.6\%\\\hline
 'Atrial Premature'  & 2779 & 02.5\%\\\hline
 'Fusion of ventricular and normal'  & 803 & 00.7\%\\\hline
 Total  & 109446 & 100\%\\\hline
\end{tabular}
\end{table}
To visualize the signal from 187 features from the data set, figure \ref{fig:ecg type} is presented. ECG signal for each class is presented with a different class. Each signal length is 175 ms with one beat.

\subsection{Motivation to use different preprocessing to increase performance}
To reduce noise and provide a cleaner, more readily interpretable ECG, ECG preprocessing is a huge task.

Filtering is utilized in ECG signal preprocessing to achieve high speed and low latency design with fewer computing parts. Because the signal processing approach was created for remote healthcare systems, noise elimination is the primary focus.
A Chebyshev type II band-pass filtering with both low and high cutoff frequencies of 0.5 Hz and 48 Hz removes baseline wandering noise as well as disturbance from the ECG signal  \cite{ayashm2022analysis}.

ECG signal denoising is a significant pre-processing step that reduces noise and emphasizes the characteristic waves in ECG data\cite{chatterjee2020review}. In our technique, the Daubechies wavelets denoising tool is beneficial for compressing and eliminating noise from ECG data \cite{38singh2006optimal}.

The goal of segmenting the ECG signal is to detect the waves and segments, including intervals and compare them to known patterns based on their timing and morphological features \cite{aspuru2019segmentation}. A proper segmentation process helps to understand the actual class label. In our method, LPD \cite{SEGlaguna1994automatic} segmentation process is used in a recent method to improve the signal levels in the classification tasks.

ECG normalization reduces the disturbance, which is a critical step in the processing of ECG signals. The normalization method not only generates a consistent signal for later automated processing but also provides a clear visual interpretation \cite{Nji2008baseline}. Normalization process nicely handles overfitting and underfitting to make the ECG signal meaningful \cite{Nzhang2021over}\cite{Nchoi2020driver}.

ECG signals can be affected by a wide range of noise and distortions. As a result, increasing the signal-to-noise ratio (SNR) before inputting the deep learning algorithm is critical.. To preprocess ECG signals, we performed the stages outlined in the appendix section \ref{ECG data preprocessing}.  At first, ECG beats were pre-processed and generated the synthesized signal with the CGAN model.

\subsubsection{Generation of Synthetic data}
\begin{figure*}[t]
\centering
\includegraphics[width=.89\textwidth]{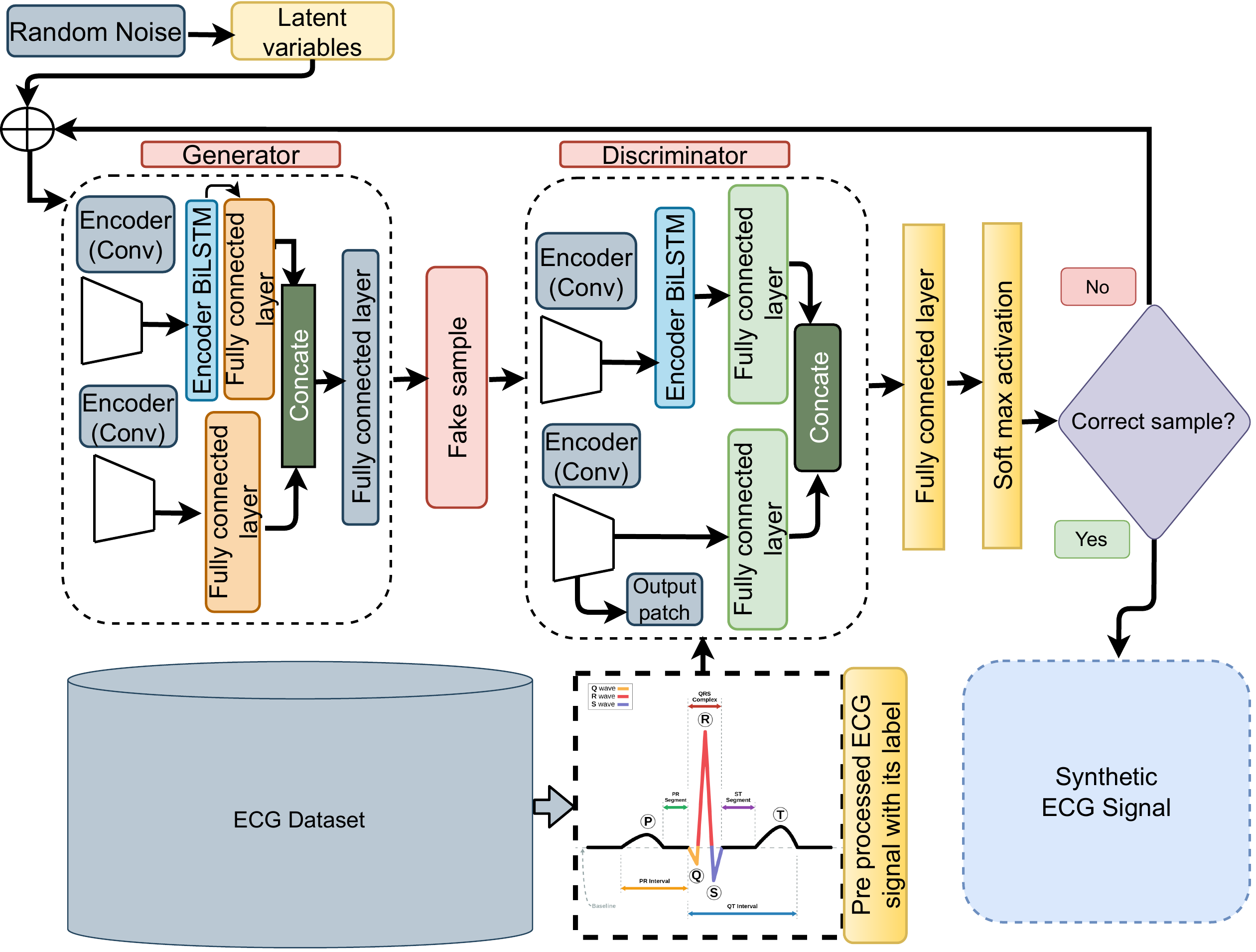}
\caption{CGAN model to generate synthetic signals.}
\label{fig:gan model}
\end{figure*}

\begin{figure*}[t]
\centering
\includegraphics[width=.81\textwidth]{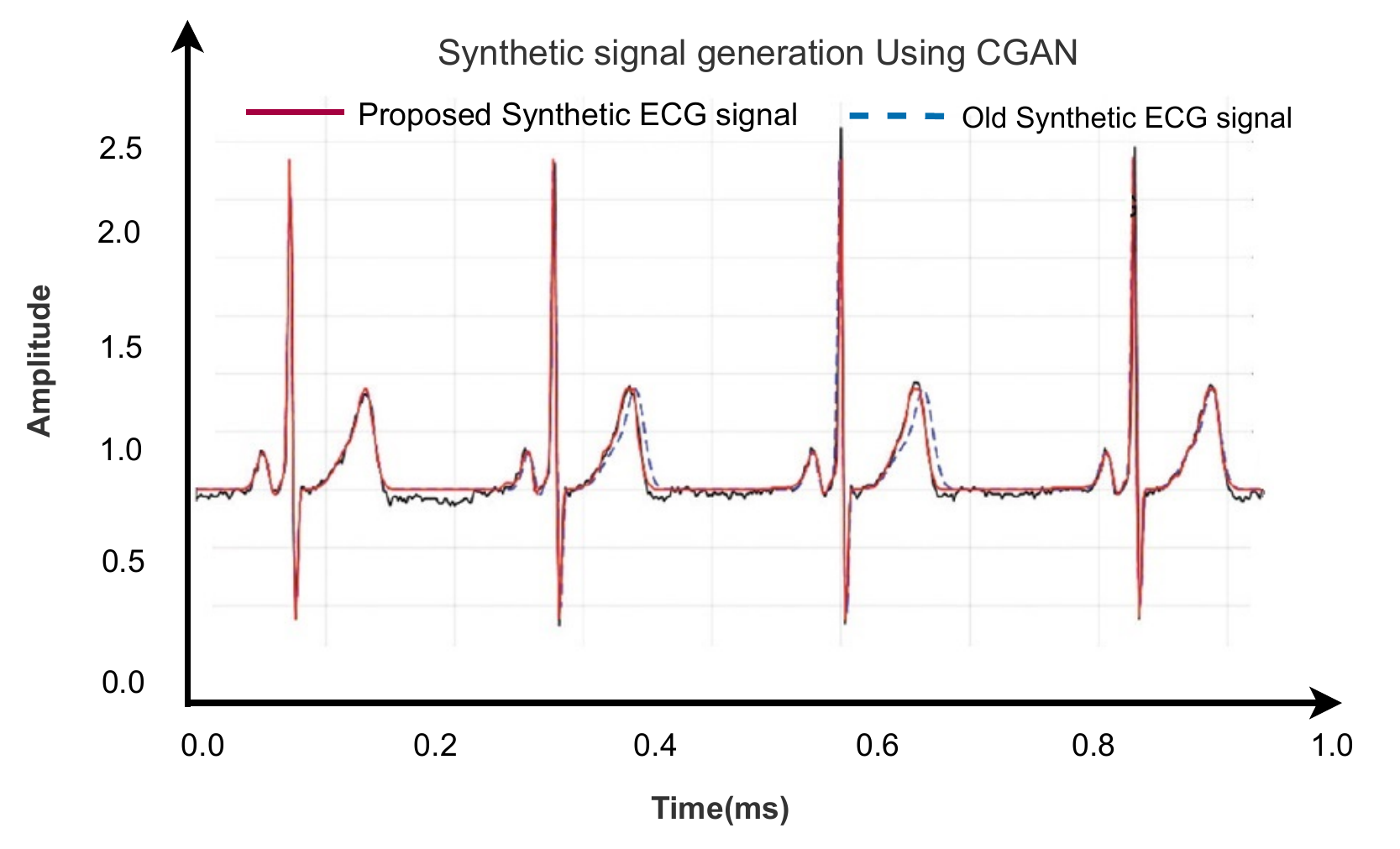}
\caption{Sample of proposed synthetic ECG generation}
\label{fig:sythetic}
\end{figure*}

GAN model is very popular in synthetic signal generation \cite{GANdelaney2019synthesis}. Synthetic data is used to compensate for the imbalance in the amount of ECG heartbeats in the five classes. After sequential preprocessing (Z Score normalization, filtering, denoising, and segmentation), synthetic data samples are generated by altering the original normalized ECG signals by our processing. Because they are the most numerous, segments in the N class remain constant. To relate segment N with raised segments. After augmentation, the gross number of segments, including five classes, has increased. The overall process of synthetic ECG signal generation is illustrated in figure \ref{fig:gan model}.
In this CGAN model, we sent a raw signal to the Discriminator part of the CGAN model, and noise with latent variable was sent to the Generator part of the CGAN model. In the generator part, encoded convolutional operational output is sent to the Encoded BILSTM layer. Fully connected layer checks are a fake or real sample, and based on the decision, its output is forwarded to the discriminator section of the CGAN model. CGAN model with Discriminator phase produces the synthetic signal after the operation of the activation function. Figure \ref{fig:sythetic} presents Raw ECG signal, and generated ECG signal. 

The confidence of the CGAN model is measured with the loss as $Loss =  E_x.[log(Dis(x))] + E_z[log(1-Dis(Gen(z))]$. Generative get real input from raw data, and an additional noise is triggered to it. Dis(x) finds a false signal and $E_x$ or $E_z$ gives the score of the reality of signal.
Dis(x) is replaced by Gen(z) in the second part of the loss function. Actually the discriminator's input is the output of Generator G when it encounters a random source tensor z. Here, generator maximise or minimise loss, whereas discriminators decrease the loss.

After all, signals are preprocessed, the synthetic signal is generated. The synthetic signal was used for classification with our HARDC model in the next step. Following this, the BiLSTM-BiGRU layer used the synthetic signal for feature extraction, and then it was sent to the next layer.

\section{Methodology}
This section discusses these connected methods by introducing a multilabel ECG signal classification methodology, flow chart, profound CNN learning model, dual-structured RNN, a layer of hierarchical attention, and other experimental factors. Here figure \ref{fig:flochar overall}shows the overall flowchart of the HARDC model. At first, the sequential preprocessing (Z Score normalize-
tion, filtering, denoising, and segmentation) and then the synthetic
data samples generation using CGAN.

A multilabel categorization is an approach that includes more than two categories. In this categorization, every label is not exclusive. The classification process determines only one degree for each sample. ECG signal labeling into one or more types is one kind of application of multilabel categorization of heartbeats in reality. We concentrated on ECG signal multilabel classification. On D data sets, We did a multiclass task as seen in equation (\ref{(4)}):. See I of this technical section for more details on this data gathering. ECG signal features and each declaration set on the vector stage are included in the data.
\begin{equation}\label {(4)}
D= (F,E)|F ∈ Documents,E ∈ (0,1)^L
\end{equation}

Where F is the ECG-based features, dataset E is an N (Number of ECG class labels) target class, and the ECG categories indicated are increased level L. 
\subsection{Main classification model}
Our HARDC approach is a hybrid profound learning method. A new hierarchical attention network with a two-way recurrent neural network and dilated convolutional neural network (CNN) is used with multi-levels in our ECG signal categorization technique. Our method has the following steps to be operated. 
\begin{enumerate}
    \item Initially, the benchmark data was gathered, and then preprocessing (Z Normalization, Filtering, Denoising, and segmentation) tasks were done on the raw data. 
    \item After completing all the preprocessing tasks then do the synthetic signal generation to make the balance of the signal. The synthetic signal is sent to layer BiGRU-BiLSTM for feature extraction.
    \item The BiGRU-BiLSTM creates a vector routing algorithm for extracting long-ranged feature information for the deep layer of the dilated CNN. 
    \item Dilated CNN does a faster feature extraction operation to get relevant features.
    \item The attention layer then receives the Dilated CNN features to handle the most pertinent ECG information, and the outcome is transferred to the next layer for predictions.
    \item Evaluation of the model on test data with different metrics score.
\end{enumerate}

Our proposed model used categorical cross-entropy, Adam optimizer, and ROC assessment approach. A multilabel ECG identification module incorporates these layers. The overall sequential process flow chart of our model is shown in figure \ref{fig:operaion flow main}.

\subsubsection{BiGRU-BiLSTM Layer}
\begin{figure*}[t]
\centering
\includegraphics[width=1\textwidth]{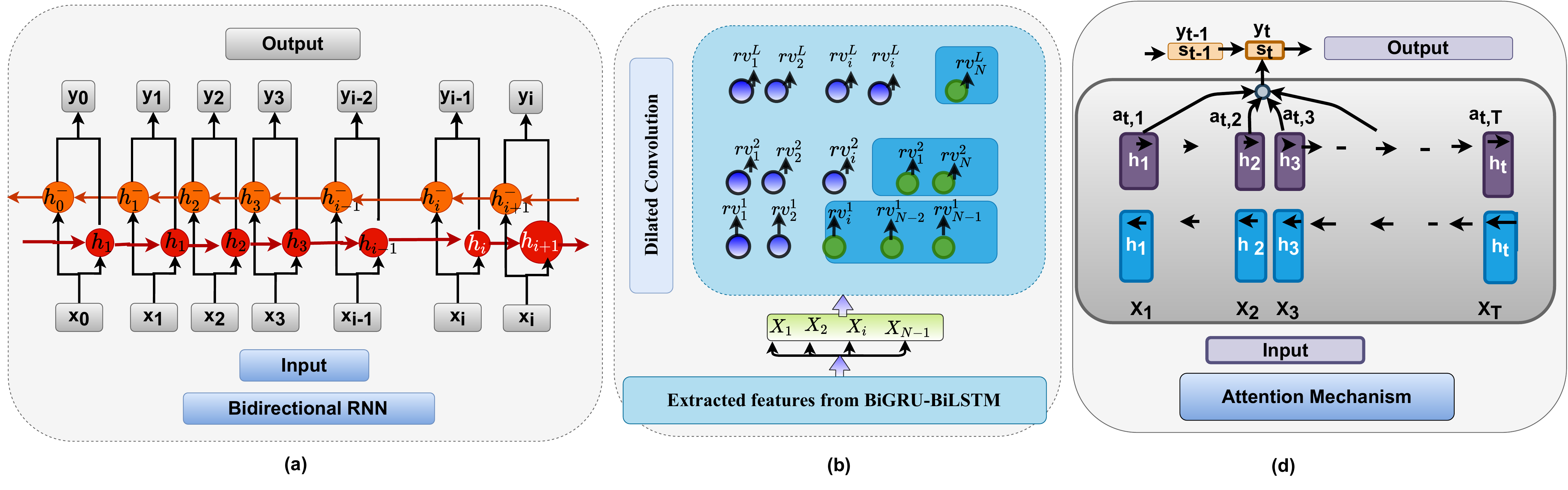}
\caption{Layer overview (a) Bi-directional RNN \\(b) Hierarchical Attention Mechanism, and (c) Dilated CNN.}
\label{fig:rnn-atten-dcnn}
\end{figure*}

The BIGRU-BiSTM layers of our model are discussed briefly in the following sections. In our method, we use two-way GRU and LSTM. This allows GRU to the functionality for long period and backpropagates through a few constrained nonlinearities, lowering the likelihood of the gradient being lost. Bidirectional setups can change the context of the 'left' and 'right' for frequent patterns. This BiGRU has two blocks, with 64 units and 128 units in the first and second, respectively. In this layer, we reduced the normal drop to 0.5 and returned the sequence as true. 
Bidirectional LSTM-GRU is frequently used to categorize ECG signals. Data is transferred from back to front in the case of a one-way LSTM-GRU. In bidirectional LSTM-GRU, data flows backward and forwards automatically thanks to two hidden states. As a result, Bi-LSTMs seem to be more familiar with the situation \cite{4dang2019novel}. BiLSTM \cite{33yildirim2019new} was used to scale the chunk of network-useable data. The basic operational sequence of Bidirectional RNN is shown in figure \ref{fig:rnn-atten-dcnn}(a).

BiGRU sequences are as below in equation (\ref{(5)}) to (\ref{(7)}):
\begin{align}\tiny \label {(5)} 
\overrightarrow{bg_{t}}=(\overrightarrow{GRU_{t}}*\overrightarrow{h_{t-1}}).x_{t}
\end{align}
\begin{align}\label {(6)}
\overleftarrow{bg_{t}}=(\overleftarrow{GRU_{t}}*\overleftarrow{h_{t-1}}).x_{t}
\end{align}
\begin{align}\label {(7)}
{bg_{t}}=(\overrightarrow{h_{t}}*\overleftarrow{h_{t}})
\end{align}

BiLSTM sequences are operated as follows in equation (\ref{(8)}) to (\ref{(10)})::
\begin{align}\label {(8)}
\overrightarrow{bl_{t}}=(\overrightarrow{LSTM_{t}}*\overrightarrow{h_{t-1}}).x_{t}
\end{align}
\begin{align}\label {(9)}
\overleftarrow{bl_{t}}=(\overleftarrow{LSTM_{t}}*\overleftarrow{h_{t-1}}).x_{t}
\end{align}
\begin{align}\label {(10)}
{bl_{t}}=(\overrightarrow{h_{t}}*\overleftarrow{h_{t}})
\end{align}

As a result, the BiGRU and BiLSTM layer encoders produce a sequence of variables as follows in equation (\ref{(11)}) to (\ref{(12)})::
\begin{align}\label {(11)}
BG= bg_{1},bg_{2},....,bg_{n}\in R^{n*d^\sim}
\end{align}
\begin{align}\label {(12)}
BL= bl_{1},bl_{2},....,bl_{n}\in R^{n*d^\sim}
\end{align}

BiGRU and BiLSTM outputs from parallel execution are combined and passed to the dilated convolution operation

This layer handles long-ranged features of ECG signal from the bidirectional operation of BiLSTM-BiGRU. 
\subsubsection{Deep CNN layer}
This is one of the most important levels in the HARDC model. The convolution layer's deep depth aims to discover hierarchical, granular qualities as architectural content descriptions. Quickly extracted relevant features from the dilated CNN layer obtained in this layer. Unlike normal CNNs, which perform convolution operations directly on pre-trained weight, our Deep CNN receives the outputs of the the Bi-GRU and Bi-LSTM layers that contains spatial information. Actually, BiGRU and BiLSTM were concatenated and sent to this Dilated CNN layer to obtain RV output. We describe this as follows for the few blocks of convolution depicted in figure \ref{fig:rnn-atten-dcnn}(b). Dilated gets concatenated features of BiLSTM and BiGRU as given in equation (\ref{(13)}) to (\ref{(15)}):
\begin{align}\label {(13)}
RV_{BiGRU}=BG= bg_{1},bg_{2},...,bg_{n}\in R^{n*d^\sim} 
\end{align}
\begin{align}\label {(14)}
RV_{BiLSTM}=BL= bl_{1},bl_{2},..,bl_{n}\in R^{n*d^\sim}
\end{align}
\begin{align}\label {(15)}
RV_l=concatenation(RV_{BiGRU},RV_{BiLSTM})
\end{align}

The scalar component d of the input's input sequence. The outputs can thus be thought of as the final pieces of each intermediary architecture.
\begin{align}\label {(16)}
RV^1=[rv^l_1,rv^l_2,....,rv^l_n]\in R^{n*d^\sim},l\in{(1,L)}
\end{align}

The overall number of blocks convolution is L, and the degree of block filters is k. Let's have a look just at block l-th amount.
\begin{align}\label {(17)}
W^l\in R^{k*w*k},W^l\in R^{k*w*d^\sim}
\end{align}
Just one k filter kernel and the w vectors are used in this filtration matrix. The following is how the two adjoining blocks can be adjusted.
\begin{align}\label {(18)}
RV=F(W^1,RV^{l-1})
\end{align}

It's basically sliding filters applied to the w-length inputs as window, where f stands for algebraic function. Formally,  $rv^l_1\in{RV^l}$ is calculated as follows 
\begin{align}\label {(19)}
rv^l_t=ReLU(W^l\oplus[rv^{l-1}_{t+1r}]^{w-1}_{t=0}
\end{align}

The $\oplus$ sign denotes the process of convolution. With CNN, r is utilized to describe the amount of deep layer. The activation function Rectified linear units (ReLU) is essentially a rectification function. The BiGRU-BiLSTM technique is a deep layer technique in which each block is increased by a maximum layer of $2^{L-2}$, and all block fields have a maximum width of $(w-1)2^{L-1}$. This deep layer is the feature we use. A typical layer of a deep learning model raises its weights exponentially rather than increasing the corresponding field weight linearly with network depth. The formula $(w-1)*2^{L-1}$ can calculate the receiving field. Finally, hierarchical maps of $RV^1, RV^2,…….,RV^l$ are retrieved. The CNN result is produced using Vector Routing, and the convolution initiative is $rv^N_l$. The dynamic approach predicts the dilated convolutional output as value $V_io$, the coefficient of connection that often changes. Both the upstream and downstream layer convolutions have a single set of coupling coefficients. SoftMax is used to calculate it using a zero bio set. $CV_{io}$'s larger one aids in correct categorization.
Currently $RV^1=[rv^l_1,rv^l_2,....,rv^l_n]\in R^{n*k^\sim},l\in{(1,L)}$
The outcome of the l-th convolutional block is represented as $R^{n*k^\sim},l\in{(1,L)}$. Each rv represents an n-gram feature generated by the k filter mapping, which we then use as input convolution. Take $CV^1=[CV^l_1,CV^l_2,....,CV^l_n]\in R^{M*d_V}$ now. The convolution's objective is dv, where M denotes the quantity of target convolution and dv denotes the output convolution size. The goal is to route $RV^l$ to $CV^l$ for increasing feature extraction interpretability and information categorizing. The predicting vector $\widetilde{rv}_{J|l}$ is used to examine raw vectors to transfer and is created by multiplying $rv_i$ by the transformation matrix $W_j$.
\begin{align}\label {(20)}
\widetilde{rv}_{J|l}=rv_i*W_j
\end{align}

This strategy increases the management and information exchange efficiency of the complex routing approach by extending little vectors and lowering big vectors into unit vectors. We used an iteratively hierarchical routing scheme over two convolution layers to calculate the intermediate state. This hierarchical attention mechanism increases the feature extraction interpretability based on the long-ranged independent features from the BiLSTM-BiGRU as well as the quickly extracted relevant feature from the dialted CNN layer. This helps to increase the ECG signal classification accuracy in prediction. In the softmax routing function, $b_{ij}$ is set to zero and modified with the $a_{ij}$ scale contract as given in equation (\ref{(21)}) to (\ref{(22)}). The size of the samples determines its $a_{ij}$ agreement. 
\begin{align}\label {(21)}
a_{ij}=cv_{i}*\widetilde{rv}_{J|l}
\end{align}
\begin{align}\label {(22)}
b_{ij}=b_{ij}+a_{ij}
\end{align}

Typically, the target convolution M indicates the number of text categories having orphan categories. Its system may be able to distinguish "background" data, such as stop words, but not task-related words. The orphan class method eliminates class overlaps and enables more efficient and scalable convolution routing. Algorithm 1 describes the hierarchical routing scheme. Each autonomous target convolution layer generates  $CV^1=[CV^l_1,CV^l_2,....,CV^l_n]\in R^{M*d_V}$ for the l-th coevolutionary block of the functional. In a unified CV, we combine the results of goal convolution.
\begin{align}\label {(23)}
CV^1=[CV^l,CV^l,....,CV^l]
\end{align}
After the operation is completed, it will be passed across the hierarchical attention layer. 
\subsubsection{Hierarchical attention layer}
Through all the adoption of each target convolution as input, the goal of this layer is to deliver a specified and aggregation actual variable. For each target convolution $cv_i\in R^{dv}$ in CV, we evaluate attention I, which reflects its relevance and contributions to the classification objective. Figure \ref{fig:rnn-atten-dcnn}(c) depicts the basic working of the attention mechanism. The attention task is computed as given in equation (\ref{(24)}) to (\ref{(25)})
\begin{align}\label {(24)}
e_i=a(q,cv_i)
\end{align}
\begin{align}\label {(25)}
a_{ij}= \frac {exp(e_{i})} {\sum\limits_{k}{exp(e_{k})}}
\end{align}

The possibility of convolution pool CV in the entire pool is shown in which q is a training program pattern vector, and k is the probability of convolution pool CV in the entire pool. The weighted total is then transferred to the downstream classification of overall target convolutions, yielding a fixed-length attention aggregation variable. Figure \ref{fig:rnn-atten-dcnn}(c) depicts a basic working diagram of the attention mechanism.
\begin{align}\label {(26)}
o=\sum\limits_{i} a_{ij}*cv_i)
\end{align}

\subsubsection{Prediction Layer}
This layer's purpose is to simply calculate the probability distribution by p(y|S), here y is the classification target. The vector o is given to the multi-layer classification via the softmax function for the fixed-length and care-oriented aggregates.
\begin{align}\label {(27)}
o=\sum\limits_{i} a_{k} cvo_i)
\end{align}

\begin{figure*}[t]
\centering
\includegraphics[width=1\textwidth]{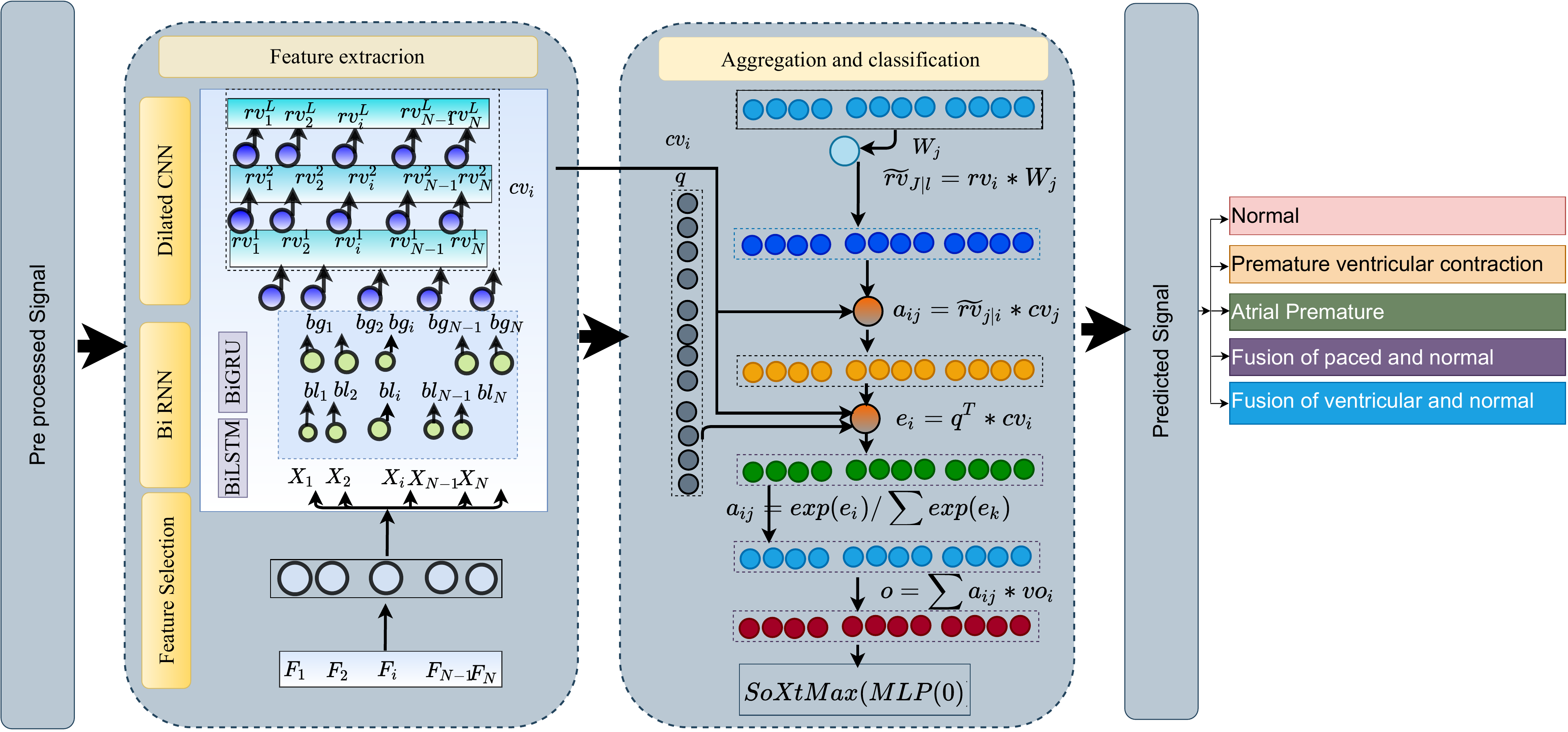}
\caption{Overall process of HARDC model.}
\label{fig:operaion flow main}
\end{figure*}

The operational phase of our model is depicted in Figure \ref{fig:operaion flow main}. It illustrates the mechanics of each layer, starting with the input section to the prediction layer. The generated output passes through the following processing layer input in one direction. We use ECG data to test our model performance in the real world. At first, the raw signal is transformed to synthetic using our CGAN model after the ECG signal is preprocessed. After filtering, denoising, and Z normalization, the signal is transformed into the deep learning model to be executed sequentially. At the final stage, the ECG target class is predicted.
\subsection{Main pseudocode of HARDC model to classify ECG signal.}
The core algorithm operation on ECG signal data is described in this section. The algorithm receives raw data as input and predicts ECG target classes based on the data. The ECG signal is initially recorded and preprocessed. The BiGRU- BiLSTM layer receives the preprocessed signal and extracts useful features. The output of the dilated CNN is transmitted to the hierarchical attention mechanism. At the last layer, SoftMax is used to forecast ECG signals. Here, Pseudocode \ref{psudocode} is the main algorithm that is used for ECG classification with HARDC model\\

\hrule
Psudocode1(): \label{psudocode}
\hrule
\KwData{Input Datasets of ECG data D, Feature list Fi}
\KwResult{Prediction of Arrhythmia}
Input synthetic signal from $D_s$\\
Input features ( F1,F2,….,FN) from the synthetic signal from ECG data\\
Pre processed features is generated as $X= x_{1},x_{2},....,x_{n}$\\
Process ECG features inputted: :$X= x_{1},x_{2},....,x_{n}\in R^{n*d}$       ,here d is dimension\\
Process generated output with BiGRU and BiLSTM                                    \\
   \For{every repetition i in range 0 to N}
   {
   $RV_{BiGRU}=BG= bg_{1},bg_{2},...,bg_{n}\in R^{n*d^\sim} $ \\
   $RV_{BiLSTM}=BL= bl_{1},bl_{2},..,bl_{n}\in R^{n*d^\sim} $ \\
   Here RV is Routed 
   Process BiLSTM-BiGRU layers:
   $RV_l=concatenation(RV_{BiGRU},RV_{BiLSTM})$\\
   $RV^l=[rv^l_1,rv^l_2,....,rv^l_n]\in R^{n*d^\sim},l\in{(1,L)} $\\
   $rv^l_t=ReLU(W^l\oplus[rv^{l-1}_{t+1r}]^{w-1}_{t=0}$\\
   Here L is the number of layers
   $W^l\in R^{k*w*k}, W^l\in R^{k*w*d^\sim}$  
   }
Execute dilated CNN \\
   \For{every repetition of l from 0 to l}
   {
      \For{every repetition of r from 0 to N}
      {
       $RV^1=[rv^l_1,rhv^l_2,....,rv^l_n]\in  R^{n*d^\sim},l\in{(1,L)} $\\
       \For{w in range(0,w)}
       {
       $rv^l_t=ReLU(W^l\oplus[rv^{l-1}_{t+1r}]^{w-1}_{t=0}$\\
       }
      }
   }

Process Dynamic convolution network on Deep CNN output:\\ 
 \For{every repetition of i from 0 to N}
  {
   \For{every repetition of j from 0 to N}
     {
     $cv_j=\sum_{i=1}^m C_{ij}*\widetilde{rv}_{J|l}$\\
     $\widetilde{rv}_{J|l}=rv_i*W_j$\\
     $a_{ij}=cvo_{i}*\widetilde{rhv}_{J|l}$\\
     }    
  }
Execute hierarchical attention mechanism\\
 \For{every repetition of i from 0 to N}
  {
   \For{every repetition of j from 0 to N}
    {
     \For{every repetition of k from 0 to N}
      {
       $e_i=a(q,cv_i)$\\
       $a_{ij}= \frac {exp(e_{i})} {\sum\limits_{k}{exp(e_{k})} }$ \\
       $a(q,cv_i)=q^Tcv_i$\\
      }
   }
  }
Execute prediction function\\
\For{every repetition of i from 0 to N}
  {
   \For{every repetition of k from 0 to N}
   {
   $o=\sum\limits_{i} a_{k} cv_i)$ 
   }
  }
Execute prediction function\\
$p(y|S) = p(y|o) = SoftMax (MLP(o))$\\
\hrule
\subsection{Evaluation Metrics} The performance is evaluated using several performance metrics that are popularly used. Some of the evaluation metrics are presented below with their computed equation. In the calculation, P indicates Positive, T indicates True, F indicates False, and N indicates Negative. In loss function y is predicted value $y_i$. Their calculations are computed as follows in the equations (\ref{(28)}) to (\ref{(33)}).
\begin{align}\tiny \label {(28)} 
Accuracy = \frac{(TP+TN)}{TP+TN+FP+FN}
\end{align}
\begin{align}\label {(29)}
Precision = \frac{(TP)}{TP+FP}
\end{align}
\begin{align}\label {(30)}
Recall/Sensitivity = \frac{(TP)}{TP+FN}
\end{align}
\begin{align}\label {(31)}
Specificity = \frac{(TN)}{TN+FP}
\end{align}
\begin{align}\label {(32)}
F1 = 2*\frac{(precision*Recall)}{precision+Recall}
\end{align}
\begin{align}\label {(33)}
Loss(y)=y_i log(\bar y_i)+(1-y_i)+log (1-\bar y_i)
\end{align}

\subsubsection{Experimental Setup}
The proposed method used BiLSTM-BiGRU dilated CNN with hierarchical attention networks. To evaluate the effectiveness of this proposed model, in our experiments, we fine-tuned the model. We applied a categorical cross-validation approach to evaluate the model. In the analytical analysis, we split the dataset into 80\% training and 20\% for validation. Note that we used the 20\% unlabelled data for testing that was from the same dataset. At each round of training,  categorical  cross-validation is used in the training time of the model. The remaining 20\% was used for validation purposes in evaluating the model. Finally, we combined all the evaluation results. The models were trained with a maximum of 100 epochs, and the batch amount is size 32. The Adam optimizer with just a learning rate of $1e^{-3}$ was employed to minimize the loss, L. To mitigate the effect of overfitting while training, we used an L2 regularisation with a coefficient of $1e^{-3}$ and a dropout approach with a probability of dropping of 0.5. The number of layers for the BiRNN, Dilated CNN, and attention was progressively configured for analyzing ECG signals from the MIT-BIH Cardiac arrhythmia database. We used the Python programming language and the Google TensorFlow deep learning framework to implement the model. We used a computer with 5 CPUs (Intel(R) 3.60 GHz), RAM with 32 GB memory, and Windows 10 operating system to execute the  categorical cross-validation. The best results were provided in the outcome section of all experiments. The proposed model's overview is described in table \ref{table parameter model} of the section \ref{Model parameter}.

\section{Result analysis}
This section describes a qualitative and comparative analysis of the results. The training set performed somewhat better than the validation set as expected, and the algorithm accumulated to a steady value, showing that the settings used to train the model are not excessive. Inside the validation model, the suggested technique obtained stable classification performance with good accuracy.
\subsection{Qualitative analysis}
\begin{figure*}[t]
\centering
\includegraphics[width=1\textwidth]{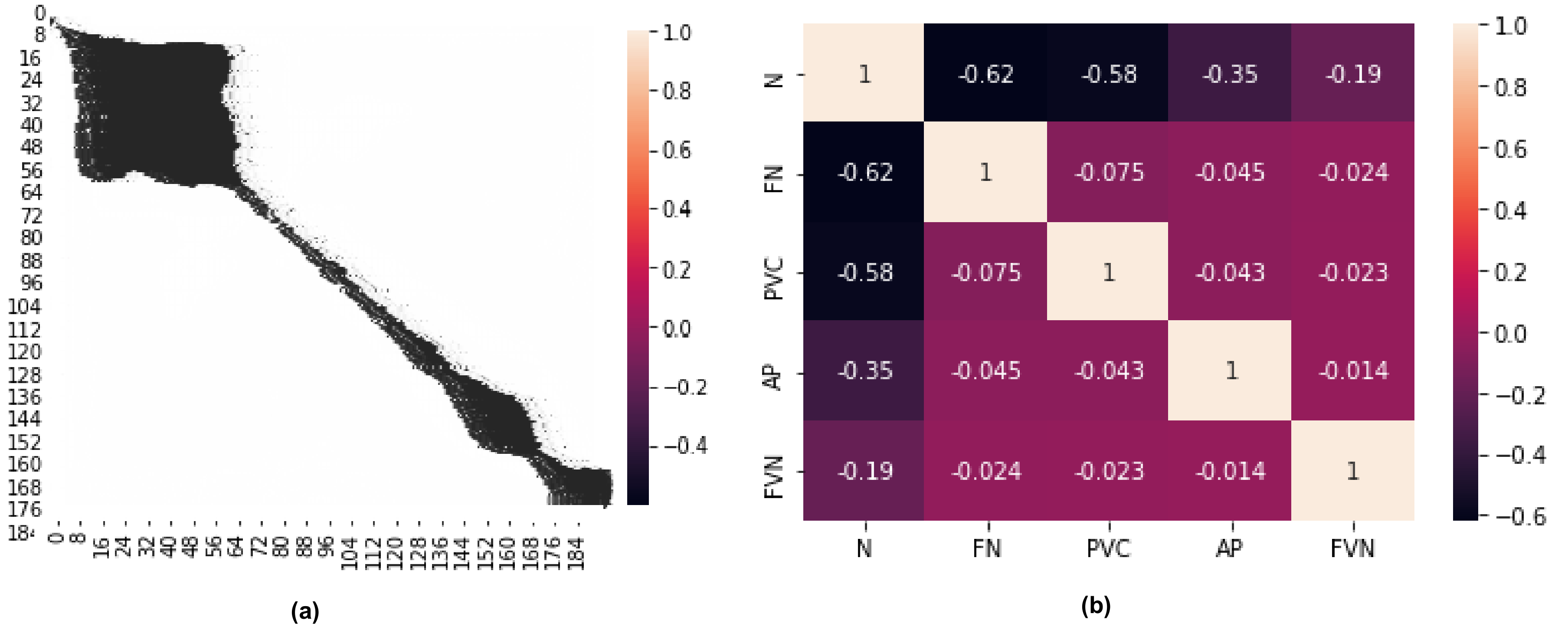}
\caption{(a) Correlation matrix among input features, \\(b) Correlation matrix among target class of ECG signal}
\label{fig:correlation matrix}
\end{figure*}

In this section, the 'Correlation Matrix' shows correlation coefficients among input or target variables. For each cell of the table, the correlation coefficient is displayed. A correlation matrix can summarize data for a more comprehensive analysis or a more advanced diagnostic tool. Figure \ref{fig:correlation matrix}(a). shows the ECG Data correlation matrix for all train features of the ECG signal. Figure\ref{fig:correlation matrix}(b). shows the ECG Data correlation matrix for the target class of ECG signal. In figure \ref{fig:correlation matrix}.b, "Normal", "Fusion of paced and normal", "Premature ventricular contraction", "Atrial Premature", and "Fusion of ventricular and normal" of ECG target class are named N, FN, PVC, AP, and FVN respectively.
In order to check the performance of our model in training and validation, we set fine parameters for our model. Batch size 32, Learning rate 1e-3, and epoch size 100 are set for the proposed HARDC model. Our method obtained an accuracy of  99.60\% for training and 99.40\% for validation in the learning time. The model also achieved a precision of 97.66\% in training and 95.69\% for validation, F1 of 98.21\% in training, and 96.56\% in validation, this is stated in the table \ref{table compared conventional model}. This HARDC model takes two hours to train. The loss function value is 0.0.0150 for training and 0.0.0440 for validation. 

\begin{table}
\caption{performance of HARDC Model}
\small
\label{table train validation }
\begin{tabular}{ |p{0.3cm}|p{3.3cm}|p{1.2cm}|}

 \hline
 \multicolumn{3}{|c|}{Performance of HARDC model} \\
 \hline
 \parbox[t]{2mm}{\multirow{10}{*}{\rotatebox[origin=c]{90}{HARDC metrics and parameters}}} & Training Accuracy & 99.60\%\\ [1ex]\cline{2-3}
& Validation accuracy & 99.40\%\\ \cline{2-3}

& Training F1 & 98.21\%\\ \cline{2-3}
& Validation F1 & 96.56\%\\ \cline{2-3}

& Training Recall & 99.60\%\\ \cline{2-3}
& Validation Recall & 98.99\%\\\cline{2-3}

& Training precision & 97.66\%\\ \cline{2-3}
& Validation precision & 95.69\%\\\cline{2-3}

& Training Loss & 0.0150\\ \cline{2-3}
& Validation Loss & 0.0440\\ \cline{2-3}

& Training Time & 2 Hours\\ \cline{2-3}
& Validation Time & 2 Hours\\ \cline{2-3}

& Batch Size & 32\\ \cline{2-3}
& Learning rate & $1e^{-3}$\\ \cline{2-3}

\hline
\end{tabular}
\end{table}
To check the testing data performance of our proposed HARDC model, We use 20\% of unlabelled data for evaluation. Testing performance is also good as training and validation. our HARDC model gave 99.01\% accuracy on test data. Kernel size 5, RNN input size 64, RNN neuron units 128, Dilated CNN Convolution 3*3, and Global dropout 0.5 was set when tuning the model. In testing performance analysis, the testing data sample was 21,886, 187. Table  \ref{table test performance} shows the testing performance of the proposed HARDC model.

\begin{table}
\caption{HARDC Model performance on test data}
\small
\label{table test performance}
\begin{tabular}{ |p{0.3cm}|p{3.3cm}|p{1.3cm}|}

\hline
\multicolumn{3}{|c|}{Testing Performance of HARDC model} \\
\hline
\parbox[t]{2mm}{\multirow{7}{*}{\rotatebox[origin=c]{90}{Performance}}} & Testing Accuracy & 99.01\%\\ [1ex]\cline{2-3}

& Testing F1 & 97.51\%\\ \cline{2-3}

& Testing Recall & 98.02\%\\\cline{2-3}

& Testing precision & 97.01\%\\\cline{2-3}

& Testing Time & 20 secs\\ \cline{2-3}
& Testing data size & 21886*187\\ \cline{2-3}

\hline
\end{tabular}
\end{table}

\begin{figure}[t]
\centering
\includegraphics[width=0.5\textwidth]{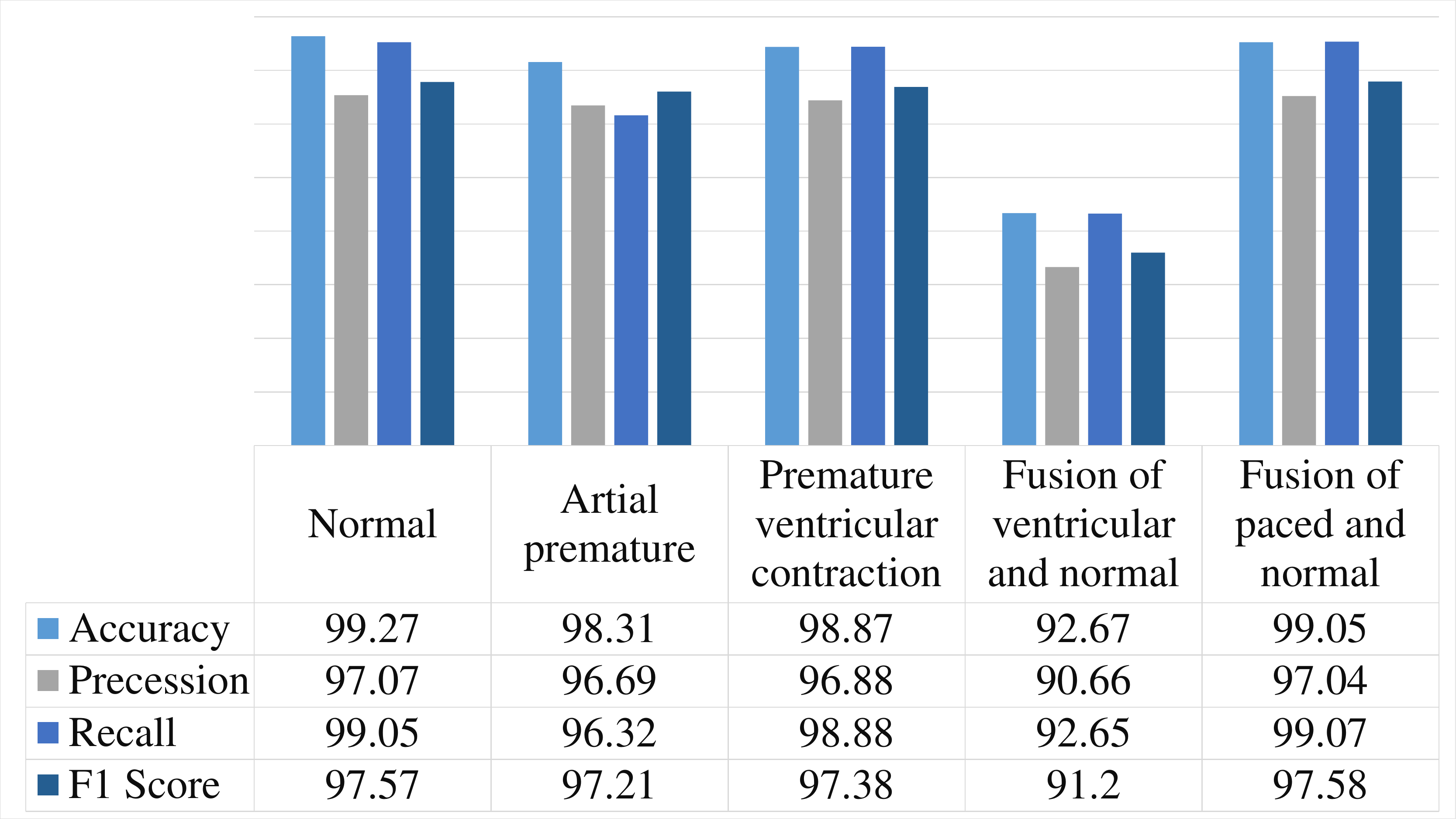}
\caption{Performance of HARDC in testing}
\label{fig: test bar}
\end{figure}

Here, figure \ref{fig: test bar} illustrates the overall testing performance of our HARDC model to classify ECG signals. In the performance evaluation, we find accuracy, F1 score, precision, and Recall. Different colours in these figures label each evaluation metric, and each bar on each subfigure presents each ECG class performance.

\begin{figure*}[t]
\centering

\includegraphics[width=.63\textwidth]{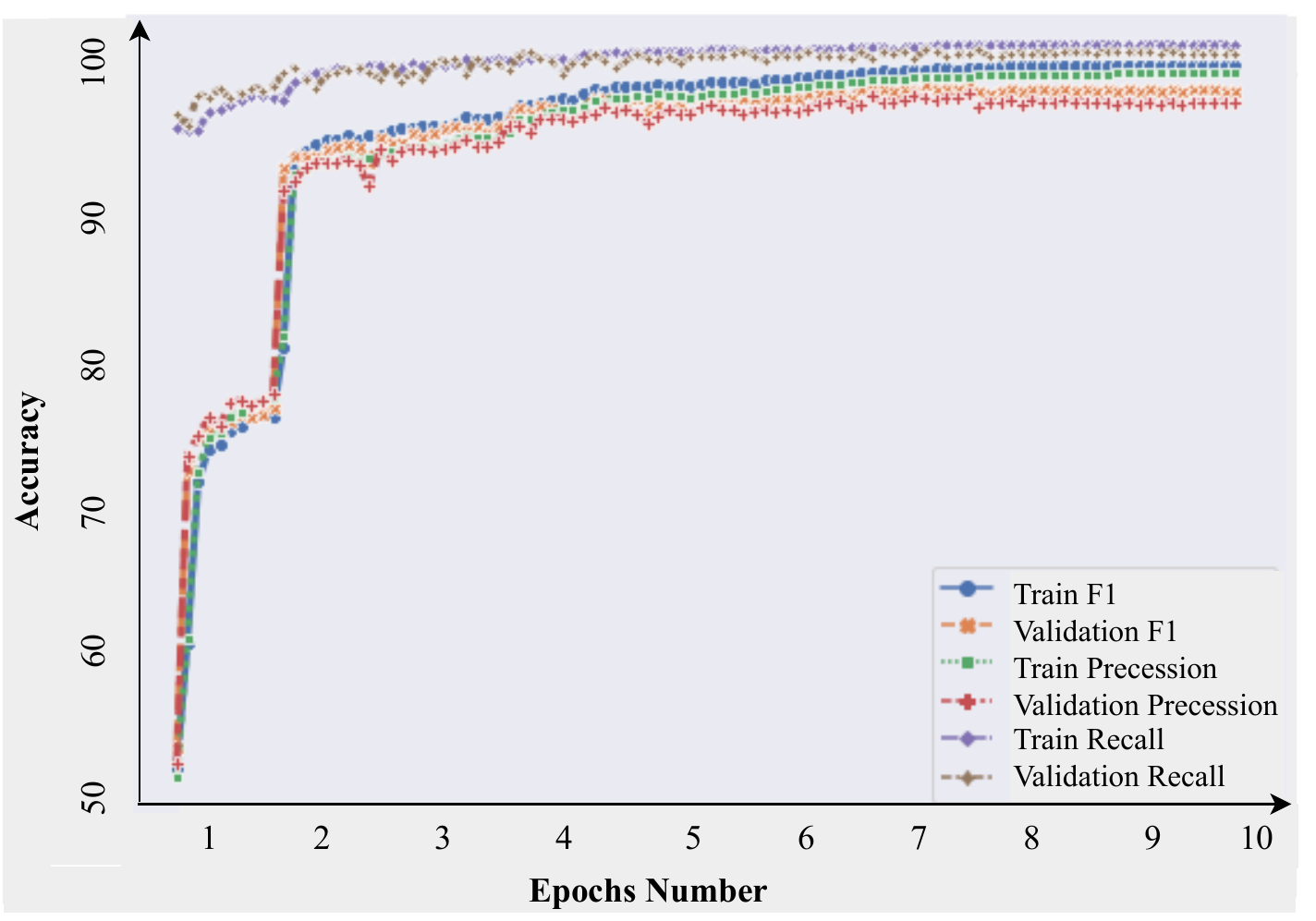}
\caption{F1 score, precision, and recall score of\\ HARDC model on ECG training data}
\label{fig:f1 p r }
\end{figure*}

Based on the training, validation, and testing data performance. Our implemented program of the HARDC model generates the following graphical result on ECG data. Figure \ref{fig:f1 p r } shows the performance during training and validation. In this figure  \ref{fig:f1 p r } , the x-axis is used for epochs, and the y-axis is used for performance percentage. Each graphical line shows the performance of our HARDC model.

Here, the categorical Cross-Entropy Loss function of our HARDC model during training and validation. We received a training loss of 0.015 and a validation loss of 0.0440.

\subsection{Comparison with state-of-the-art technique }
This section will cover all state-of-the-art methods and our proposed method performance to classify ECG data from MIT-BIH database. All compared methods in table  \ref{table main} are classified into three categories: traditional machine learning(ML), deep learning(DL), and hybrid deep learning(HDL). Most popular traditional machine learning  algorithm in ECG signal classification SVM: \cite{21ye2012heartbeat}\cite{22chen2017heartbeat} \cite{26martis2012application}\cite{23mar2011optimization}\cite{25mondejar2019heartbeat} \cite{27elhaj2016arrhythmia} \cite{28yang2018automatic}. XGBoost \cite{24shi2019hierarchical},  Random Forest \cite{29li2016ecg}, and LDA-MLP \cite{23mar2011optimization} methods also show their performance in ECG classification in recent times. SVM classification with Wavelet coefficient and RR interval ECG signal classification \cite{21ye2012heartbeat} with 89.02\% accuracy, 62.2\% precision, and 56.27\% recall. 

Another SVM \cite{22chen2017heartbeat} classifier with RR and ECG projection matrix classified variable-length ECG signals with 98.46\% accuracy. 
LDA-MLP \cite{23mar2011optimization} method with temporal and morphological features to classify ECG signal with 89.02\% accuracy, XGBoost with handcraft features \cite{24shi2019hierarchical} for ECG classification got 91.87\% accuracy. RR intervals, Wavelet coefficient, and Morphological descriptor used in SVM-based ECG classification \cite{25mondejar2019heartbeat}. All other SVM-based classifiers \cite{28yang2018automatic} \cite{25mondejar2019heartbeat} \cite{21ye2012heartbeat} to classify ECG signals got accuracy between 91\% to 97\%. It is observed that the SVM \cite{27elhaj2016arrhythmia} classifier with multiple features to classify ECG signals with the highest accuracy, 98.91\%, among traditional machine learning methods.
The most common deep learning-based method is CNN, which here clearly showed superior performance with greater facility compared to other methods.

A recently developed method employing CNN \cite{30kiranyaz2015real} determined patient-specific features with as high as 99.00\% accuracy. 
Another method using CNN \cite{31zubair2016automated} obtained 92.70\% accuracy for classifying 0.5s long ECG signal with 200 samples. Mostly this used CNN-based methods with the generation of synthetic data and rescaling of raw data to classify ECG signals with 94.03\% accuracy. A method with random projections and extracted RR-interval features and variable length of ECG signal using CNN \cite{32sellami2019robust}  with 88.33\% accuracy, 64.2\% precision, and 57.0\% recall. 
A hybrid method with the combination of CNN with RNN gave a precision of 98.10\% \cite{2oh2018automated}.
Yıldırım et al. \cite{1yildirim2018arrhythmia} proposed a CNN-LSTM based. This method used rescaling tools to preprocess ECG signals and it obtained 91.33\% accuracy in ECG classification.
\begin{landscape}
\begin{table}
\caption{Compared Performance.}
\centering
\small
\label{table main}
\begin{tabular}{|p{.6cm}|p{1cm}|p{3.1cm}|p{4.4cm}| p{.8cm}|p{1cm}|p{1cm}|p{8.3cm}|}


\hline

Type 
& Author 
& Method 
& Input 
& Class 
& Samples 
& Signal length 
& Performance\\ [1ex]
\hline

\multirow{2}{*} 
{ML}
& \cite{23mar2011optimization} 
& LDA-MLP 
& Temporal,Morphological  
& 5 
& NA 
& NA 
& Accuracy 89.02\%
F1 62.20\%
precision 56.27\% \\[1ex]

\cline{2-8}

& \cite{21ye2012heartbeat} 
& SVM 
& Temporal,Morphological  
& 5-15 
& Various
 & .083s & Accuracy 86.55\%
F1 55.90\%
precision 53.40\% \\[1ex]

\cline{2-8}

& \cite{22chen2017heartbeat}
& SVM
& RR interval, projection matrix
& 5-15
& Various
& .083s & 
Accuracy 98.46\%
F1       98.46\%
precision 98.46\% \\[1ex]

\cline{2-8}


& \cite{24shi2019hierarchical}
& XGBoost
& 168 handcrafted
& 6
& NA
& .08s
& Accuracy 91.87\%
F1       67.80\%
precision 61.80\% \\[1ex]

\cline{2-8}

& \cite{25mondejar2019heartbeat}
& SVM
& RR, Wavelet, Morphological
& 5
&260
& NA 
& Accuracy 94.4\%
F1       67.00\%
precision 66.30\% \\[1ex]

\cline{2-8}

& \cite{26martis2012application}
& LS-SVM
& Statistical PCA
& 5
&200
& 0.56s 
& Accuracy 98.11\%
precision 90.60\% \\[1ex]

\cline{2-8}

& \cite{27elhaj2016arrhythmia}
& SVM
& PCA,DWT,HOS,ICA
& 5
&200
& 0.56s 
& Accuracy 98.91\% \\[1ex]
\cline{2-8}

& \cite{28yang2018automatic}
& Linear SVM
& Wavelet coefficient
& 5
&300
& 0.83s 
& Accuracy 97.94\%
Precision       98.05\%
Sensitivity       99.56\%

Specificity 89.33\% \\[1ex]
\cline{2-8}

& \cite{29li2016ecg}
& Random Forest
& Wavelet coefficient
& 5
& 143
& 0.39s 
& Accuracy 98.30\%
Precision       79.30\%
Sensitivity       99.05\% \\[1ex]
\cline{1-8}

\multirow{2}{*} 
{DL}
& \cite{30kiranyaz2015real} 
& CNN 
& Patient-Specific   
& 5 
& 64/128 
& Variable 
& Accuracy 99.00\%
Sensitivity 79.30\%
Specificity 99.05\% \\[1ex]
\cline{2-8}

& \cite{31zubair2016automated}
& CNN
& Patient-Specific
& 5
& 200
& 0.56s 
& Accuracy 92.70\% \\[1ex]
\cline{2-8}

& \cite{13yildirim2018efficient}
& CNN
& synthetic signal, Re-scaling 
& 5
& 360
& 1s 
& Accuracy 94.03\% \\[1ex]
\cline{2-8}

& \cite{32sellami2019robust}
& CNN
& Random projections, 

RR-interval
& 5
& 260
& Variable 
& Accuracy 94.03\%
F1 64.20\%
Precision 57.00\% \\[1ex]
\cline{1-8}

\multirow{2}{*} 
{HDL}
& \cite{20waldo2008inter} 
& CNN-LSTM 
& Patient-Specific  
& 5 
& Various 
& Variable 
& Accuracy 98.10\%
Sensitivity 97.50\%
Specificity 98.70\% \\[1ex]
\cline{2-8}

& \cite{1yildirim2018arrhythmia}
& DCNN 
& Re scaling Raw data
& 17 
& Various 
& 0.015s 
& Accuracy 91.33\% \\[1ex]
\cline{2-8}

& \cite{33yildirim2019new} 
& CAE-LSTM 
& Coded signals   
& 5 
& 260 
& 0.72s 
& Accuracy 99.10\%
Sensitivity 99.00\%
Specificity 99.00\% \\[1ex]
\cline{2-8}

& \cite{5chen2020multi} 
& CNN-BiLSTM 
& RR intervals
& 5 
& 224/325 
& 1.53s 
& Accuracy 96.77\%
F1 77.80\%
Precision 81.20\% \\[1ex]
\cline{2-8}

& \cite{12ma2021ecg} 
& Dilated CNN 
& Denoising, normalization   
& 5 
& 260 
& NA 
& Accuracy 98.65\%
Sensitivity 97.98.97\%
Specificity 99.03\% \\[1ex]
\cline{2-8}

& \cite{C1tao2022ecg} 
& Deep Attention 

BiLSTM 
& Attention,morphological,

temporal   
& 5 
& 360
& Various
& precision 96.7\%
Recall 96.4\%
F1 96.5\% \\[1ex]
\cline{2-8}

& \cite{4dang2019novel} 
& Deep CNN-BiLSTM 
& RR, PR intervals, QRS   
& 5 
& 100 
& 0.8s 
& Accuracy 96.59\%
Sensitivity 99.93\%
Specificity 97.03\% \\[1ex]
\cline{2-8}

& \cite{34jin2020multi} 
& CNN-LSTM 

Attention 
& Multi domain features   
& 5 
& 360 
& NA 
& Accuracy 98.51\%
Sensitivity 98.14\%
Specificity 98.76\% \\[1ex]
\cline{2-8}

& \cite{11wang2021intelligent} 
& CNN-BiLSTM 

Attention 
& Raw data   
& 5 
& 360 
& NA 
& Accuracy 99.10\%
Precision 98.16\%
Sensitivity 98.40\%
Specificity 98.70\% \\[1ex]
\cline{2-8}

& HARDC (Our) 
& Attention 

BiGRUBiLSTM 

Dilated CNN 
& Wavelet, QRS, Synthetic
& 5 
& 360 
& 1.5s 
& Accuracy 99.60\%
F1         98.21\%
precision 97.66\%
Recall     99.60\% \\[1ex]
\cline{1-8}

\end{tabular}
\end{table}
\end{landscape}
 This author also developed a method with attention to LSTM and obtained 99.11\% precision \cite{33yildirim2019new}. Another new method using CNN and BiLSTM \cite{5chen2020multi} gave an accuracy of 96.77\%, 77.8\% precision, and 81.2\% recall obtained by this method. Dilated CNN-based method developed by Ma et al. \cite{12ma2021ecg} with utilizing data denoising and normalization to detect arrhythmia with 98.65\% accuracy. 
Dang et al. \cite{4dang2019novel}. proposed CNN with a BiLSTM-based method to classify ECG signals obtained 69.59\% precision. Jin et al \cite{34jin2020multi} proposed a technique utilizing CNN-LSTM with attention that uses domain knowledge and got 98.51\% accuracy in classifying five types of ECG signals. Another technique with CNN-BiLSTM \cite{11wang2021intelligent} obtained 99.11\% accuracy for ECG classification. Compared Performance  is presented in table  \ref{table main}. A multivariate ECG classification by residual channel attention networks \cite{C1tao2022ecg} and BiGRU obtained 96.7\% and 97.7\% accuracy for long and short-length ECG signals. This method used Lead attention and morphological and temporal features very well. This proposed HARDC method used synthetic and preprocessed ECG signals with 360 samples of 1.5s long. In comparison, our hybrid HARDC model achieved the highest accuracy of 99.60\%, F1 of 98.21\%, and 97.66\% precision.

\subsection{Discussion}
The literature section outlines the issues with existing approaches, which our new HARDC model is designed to address. The HARDC model we propose was assessed and compared to several current model benchmarks. Our HARDC model had 99.60\% accuracy, 98.21\% F1, 97.66\% precision, and 99.60\% recall on MIT-BIH provided ECG data. We synthesized raw ECG signals using the CGAN model before applying our HARDC model after the preprocessing operations. 

Based on MIT BIH ECG data, we performed five categorization procedures. Experiments revealed that our strategy outperforms several competing baselines and delivers several cutting-edge results.

With the techniques of our successive pre-processing completed for the raw normal ECG signals, the CGAN model significantly improved the generation of Synthetic data samples in our method. This aids in the resolution of misclassification issues and improves performance. 
An attention-based network with dual structured BiGRU and BiLSTM is the key aspect of our proposed HARDC model, which uses the hybrid dilated CNN approach. We propose a new hybrid model for improving learning structure, extracting features, and analyzing ECG signal categorization by dynamically changing its hierarchical system into an attentive convolution. The processing of the BiGRU-BiLSM and dilated CNN, along with a system of hierarchical self-attention, can automatically extract ECG features' hierarchical representations to exploit the features thoroughly. The implicit feature information obtained by our proposed hybrid neural network convolution model was successful. The difference and interdependence of the BiGRU-BiLSTM were used throughout this model.

\begin{table*}
\caption{Compared performance of HARDC model with conventional models.}
\label{table compared conventional model}
\small
\begin{tabular}{|p{3.9cm}|p{1cm}|p{1.5cm}|p{1.4cm}|p{1.3cm}|p{1cm}|p{1.7cm}|p{1.3cm}|}
\hline

Models & Batch amount & Learning function & Activation & Optimize function & Drop out & Input-size & Accuracy \\ [1ex]
\hline

CNN & 32 & $1e^{-3}$ & Re-Lu & Adam & 0.5 & 300.128.5 & 96.00\% \\ [1ex]
\hline
LSTM & 32 & $1e^{-3}$ & Re-Lu & Adam & 0.5 & 300.128.5 & 96.50\% \\ [1ex]
\hline
GRU & 32 & $1e^{-3}$ & Re-Lu & Adam & 0.5 & 300.128.5 & 97.84\% \\ [1ex]
\hline
Attention CNN-BiGRU & 32 & $1e^{-3}$ & Re-Lu & Adam & 0.5 & 300.128.5 & 98.43\% \\ [1ex]
\hline
Attention CNN-BiLSTM & 32 & $1e^{-3}$ & Re-Lu & Adam & 0.5 & 300.128.5 & 98.87\% \\ [1ex]
\hline
Attention DCNN-BiGRU & 32 & $1e^{-3}$ & Re-Lu & Adam & 0.5 & 300.128.5 & 98.93\% \\ [1ex]
\hline
HARDC(Our) & 32 & $1e^{-3}$ & ReLu & Adam & 0.5 & 300.128.5 & 99.60\% \\ [1ex]
\hline

\end{tabular}
\end{table*}

To improve performance, the dilated CNN has the ability to handle rich features. Using a paradigm based on attentiveness and including hierarchical self-handling methods, our hybrid model takes dilated CNN that provides a reduction in training time and a clear network structure to boost performance. The improved convolution network with dynamic routing algorithm that improves the efficacy of the extraction of more relevant features.

We have also implemented some conventional methods with the same parameter tuning and input size to evaluate the effectiveness of our proposed model. Table \ref{table compared conventional model} presents the compared performance. Based on the table  \ref{table compared conventional model}, HARDC method performs better than all other comparable methods to date.

The overall performance comparison of HARDC model with the most common and related models is shown in the table conventional model which indicates that our method was able to obtain greater accuracy than other methods with the same parameter tuning.

\subsection{Complexity Analysis}
We developed a hybrid deep learning model HARDC employing Bidirectional RNN and dilated CNN with a hierarchical attention mechanism to ensure the effectiveness of the method in consideration of running time. The model's recorded running time is faster than any other deep learning approach on the testing set. Because training a convolution neural network (CNN) takes more time, the proposed method does not utilize the general CNN. However, we used the dilated convolution operation with its multilevel dilation mechanism to perform the convolution that is needed in the extraction of features, decreasing the overall computational complexity. The testing set has a 20-second run time. Convolution kernels are also firmly coupled in typical convolution layers. However, in the suggested approach, the addition of dilated components reduces the convolutional layer's processing complexity. Our model thus also has a modest spatial complexity and takes less time than some models which used an attention mechanism with normal CNN convolution and RNN, as shown in table \ref{table test time}. Our model takes less computation time through the use of dilated CNN but takes more time than some conventional models like CNN, GRU, and LSTM. In the future development, we will focus to reduce the time complexity by using different optimisers, normalization, and regularization.  
\begin{table*}
\caption{Compared testing time.}
\label{table test time}
\small
\begin{tabular}{|c|c|c|}
\hline

Model & Sample size & Time(s)  \\ [1ex]
\hline

CNN & 21,886 & 15 \\ [1ex]
\hline
LSTM & 21,886 & 18.5  \\ [1ex]
\hline
GRU & 21,886 & 19 \\ [1ex]
\hline
Attention CNN BiGRU & 21,886 & 26 \\ [1ex]
\hline
Attention CNN BiLSTM & 21,886 & 23 \\ [1ex]
\hline
Attention Deep CNN BiGRU & 21,886 & 22  \\ [1ex]
\hline
HARDC(Our) & 21,886 & 20 \\ [1ex]
\hline

\end{tabular}
\end{table*}

\subsection{Limitation of this study}
This proposed method does have some limitations. Firstly, the proposed technique only focuses on five types of signals to detect arrhythmia. Secondly, the requirement for a data-driven deep model, our future target is to analyze big and multi-label datasets to check the proposed method and improve the model generalization capabilities. We will improve more on our model complexity and adaptability in future projects.

\section{Conclusion}
Automatic categorization, particularly arrhythmia detection, is one of the most critical clinical problems and a recognized challenge in AI-based bioinformatics. Our novel method utilized an attention-based dilated CNN-BiLSTM network to diagnose arrhythmia from five types of ECG signal classes. The model integrates the feature extraction methods with improved interpretability utilizing attention-based dilated CNN, BiLSTM, and BiGRU. Bidirectional RNN, composed of two BiLSTM and BiGRU blocks, extracts relevant and long-ranged independent features. The Dilated CNN layer is a casual convolutional layer used to extract features of the ECG signals. Finally, our proposed method achieved an accuracy of 99.60\% during training, 99.40\% during validation, and 99.01\% during testing. It obtained 98.21\% F1, 97.66\% precision, and 99.60\% recall on the training dataset. According to our experimental results, this proposed method is also successful for atrial fibrillation (AF) detection, notably achieving high detection accuracy with sensitivity. Unlike previous feature learning approaches that are commonplace in biomedical signal processing, the proposed HARDC model does not require any pre-feeding of knowledge or biological information, or the handmade feature extraction procedures employed in classic machining learning can be considered as additional benefits. As a result, the network detects AF with minimal computing overhead reflected in theoretical real-time performance. The development of a cloud-based automated method to classify multi-type arrhythmia signals will be focused on  future research aligned with this work. We intend to improve the model's generalization ability, optimize neural network academic achievement, improve clinical benefits, and have a plan to decrease the time complexity of our model by using different optimizers, normalization, and regularization.

\section{Acknowledgement}
This research is not supported by any fund or grant.



\section{Appendix}
\begin{figure*}[t]
\centering
\includegraphics[width=1\textwidth]{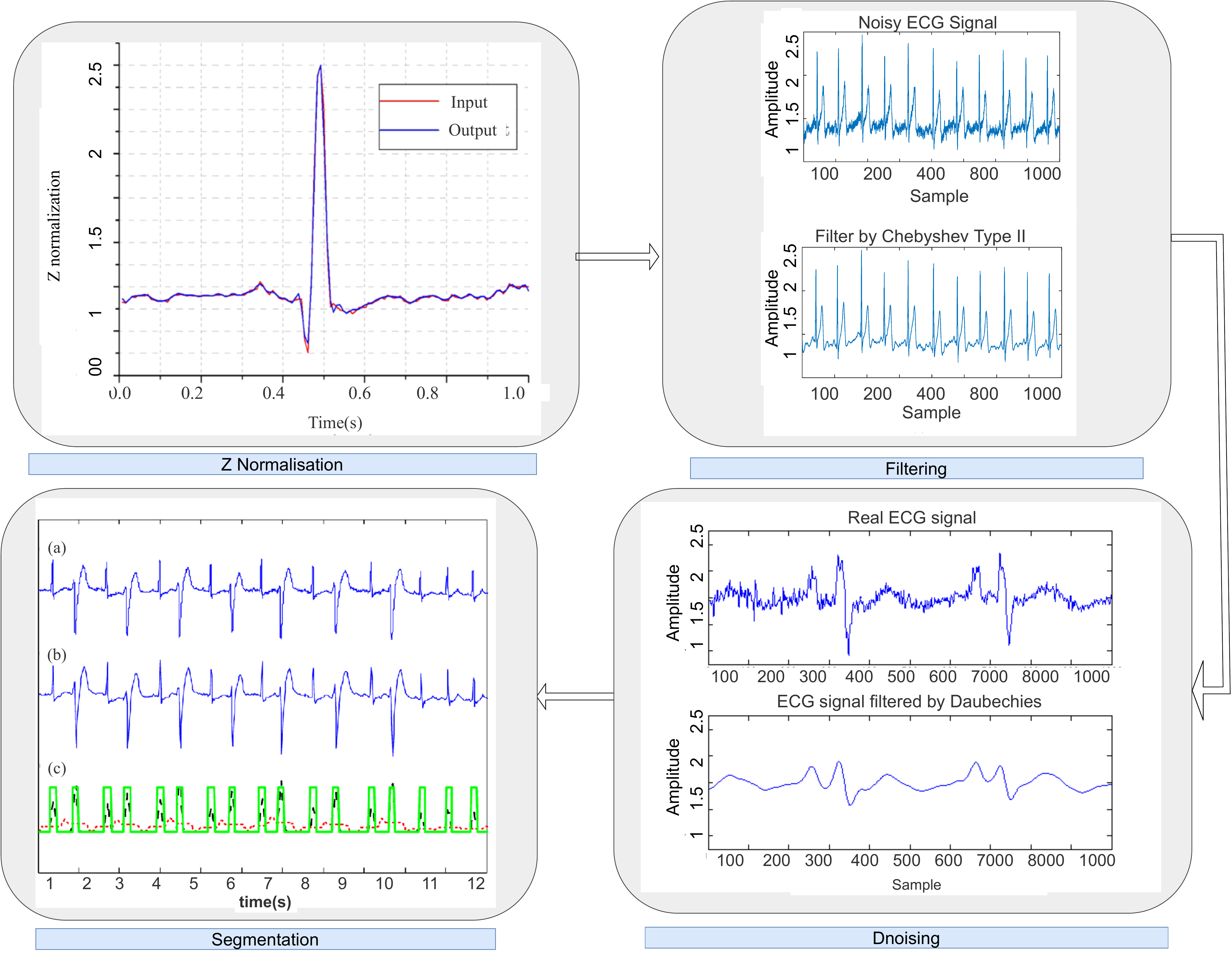}
\caption{Preprocessing of ECG signal}
\label{fig:All preprocessing}
\end{figure*}

\subsection{ECG data preprocessing} \label{ECG data preprocessing}
Numerous types of noise and artifacts can affect ECG signals. As a result, it is essential to increase the signal-to-noise ratio (SNR) before feeding to the deep learning model. To preprocess ECG signals, we performed the stages outlined below. At first, ECG beats were preprocessed and generated the synthesized signal with the CGAN model.

\subsubsection{Normalization}
R-peak detection was performed using the Pan-Tompkins \cite{36pan1985real} algorithm. All ECG data are divided into 360-sample segments centered on the observed R-peaks. Each segment was then normalized using Z-score normalization to address amplitude scaling and to reduce the offset effect before being fed into the learning model. The basic operational equation of Pan-Tompkins Z normalization is given below in equation (\ref{(1)}):
\begin{align}\tiny \label {(1)} 
ZN(z)=\frac{(x-\mu)}{\sigma}
\end{align}

Where ZN(z) denotes Z-normalization using a high pass filter.
The sequential operation of each layer is presented in this section with the execution equation (\ref{(1)}) and figures \ref{fig:All preprocessing}. 

\subsubsection{Filtering}
Noise was removed using a filtering method as the signal levels are generally so low in ECG work, thus it is vital to utilize filtering to clear a huge spectrum of noise \cite{37zarei2020performance}. We applied "Chebyshev Type II" passbands to retain the frequency band range from 0.5 to 48 Hz. This filter was chosen for its sharp cutoff frequency and lack of ripple. The primary operational equation of this filter is shown in equation (\ref{(2)}):

\begin{equation}\label {(2)}
|CH_n(j\omega)|=[\frac{1}{\sqrt{1+(∈)^2}T^2_n(\frac{\omega}{\omega_0})}]
\end{equation}
Where $∈$ is the ripple factor, and $\omega_0$ is the cutoff frequency, and $T_n$ is a Chebyshev polynomial of the nth order. Applied Chebyshev filtering is shown in the figure \ref{fig:All preprocessing}.

\subsubsection{Denoising}
After normalization, denoising was performed. ECG data was initially denoised as well as baseline removed using Daubechies wavelet filters \cite{38singh2006optimal}. Secondly, signals were split into beats and sorted according to cardiologist annotation. Basic computation \cite{DWrowe1995daubechies} of "Daubechies" wavelet filters is shown below in equation (\ref{(3)}):
\begin{equation}\label {(3)}
P_N(y) = \sum _{k=0}^{(N=1)}[\frac{N-1+k}{k}]y^k
\end{equation}
Where y is the denoised output with normalization, k indicates the segment value with a specific range. Applied Daubechies denoising is shown in the figure \ref{fig:All preprocessing}.

\subsubsection{Segmentation}
A normal rhythm generates four distinct entities \cite{ECGSEGALL-beraza2017comparative}\cite{ECGlyakhov2021system} \cite{ECG2randazzo2016ultimate}: a P wave, a QRS complex, a T wave, and a U wave, each with its own pattern. Atrial depolarization is denoted by the P wave. Ventricular depolarization is denoted by the QRS complex. Ventricular repolarization is denoted by the T wave. Papillary muscle repolarization is denoted by the U wave. The PR interval is the difference between the start of the P wave and the start of the QRS complex. The identification of QRS complexes, P and T waves, and their forms, amplitudes, and relative placements, are all part of the ECG analysis. Identifying the onsets and offsets of QRS complexes and P and T waves is known as ECG signal segmentation or delineation \cite{39azami2011novel}.

In our task, we applied Low-pass differentiation (LPD) for ECG segmentation.
LPD \cite{SEGlaguna1994automatic} is used in a recent method to improve the signal levels in the classification tasks. The lowpass differentiator produces a filtered signal that includes slope information. An adaptive threshold is used to identify QRS information of each signal. The first derivatives of the signal identify the peaks of the waves of the original ECG signal as zero crossings. Between a positive and a negative deflection, the zero-crossing represents a maximum in the initial ECG signal. The original signal wave boundaries (onset and offset) are identified by scanning backward and forward from the observed peak on the modified signal. This process performs the task as follows: I. First, detect QRS(i) and preprocess each lead $QRS_j(i)$, II. Then find missed detection if an eliminated portion is max, it can give a miss detection, and the max value can be set to the next beat. Thus, if max set to lead j, we assign $QRS(l) = QRS(l - 1) (l>= i + 1)$, here l indicates lead of ECG. 
The Applied LPD segmentation technique is shown in the figure \ref{fig:All preprocessing}.

According to the annotation offered with the dataset, all individual recordings are split into 75s segments. Each segment includes 360*75 = 27000 sample points and a corresponding label because the signals are sampled at 100Hz.

\subsection{Model parameters} \label{Model parameter}
In this subsection we have presented our model parameters on the table \ref{table parameter model}. This table presents the layer information with kernel and filter size and input size from feature extraction to prediction.
\begin{table*}
\caption{Parameters used in HARDC model.}
\label{table parameter model}
\small
\begin{tabular}{|c|c|c|c|c|}
\hline

Layer type & No. Of Layer & Layer Information

 & Filter*Kernel Size & Input Size \\ [1ex]
\hline

\multirow{8}{*} {Dilated Convolution} & 0 & 2D Convolution

(64,64,Dilation=1,Stride=1) & 64*8 & 360*128*5\\ [1ex]
\cline{2-5}
& 1 & Batch Normalization (64) & 64*8 & 360*128*5\\ [1ex]
\cline{2-5}
& 2 & ReLu(64) & 64*8 & 360*128*5\\ [1ex]
\cline{2-5}
& 3 & 2D Convolution

(64,64,Dilation=1,Stride=1) & 64*8 & 360*128*5\\ [1ex]
\cline{2-5}
& 4 & Batch Normalization (64) & 64*8 & 360*128*5\\ [1ex]
\cline{2-5}
& 5 & ReLu (64) & 64*8 & 360*128*5\\ [1ex]
\cline{2-5}
& 6 & 2D Convolution

(64,64,Dilation=1,Stride=1) & 64*8 & 360*128*5\\ [1ex]
\cline{2-5}
& 7 & Batch Normalization (64) & 64*8 & 360*128*5\\ [1ex]
\hline

\multirow{3}{*} {BiLSTM-BiGRU} & 8 & Max pooling & 64*8 & 360*128*5\\ [1ex]
\cline{2-5}
& 9 & BiLSTM(64) & 64*128 & 360*128*5\\ [1ex]
\cline{2-5}
& 10 & BiGRU (64) & 64*128 & 360*128*5\\ [1ex]
\hline

\multirow{3}{*} {Attention Mechanism} & 11 & Attention Mechanism (64) & 64*128 & 360*128*5\\ [1ex]
\cline{2-5}
& 12 & Fully Connected layer (64*128) & 64*128 & 360*128*5\\ [1ex]
\cline{2-5}
& 13 & tanh (64) & 64*128 & 360*128*5\\ [1ex]
\hline

\multirow{5}{*} {Prediction} & 14 & ReLu (64) & 64*128 & 360*128*5\\ [1ex]
\cline{2-5}
& 15 & Max Pooling (54) & 64*128 & 360*128*5\\ [1ex]
\cline{2-5}
& 16 & Global Average Pooling (64) & 64*128 & 360*128*5\\ [1ex]
\cline{2-5}
& 17 & Drop out (0.5) & Dropout = 0.5 & 360*128*5\\ [1ex]
\cline{2-5}
& 18 & Fully Connected Layer (5) & - & 5\\ [1ex]
\hline

\end{tabular}
\end{table*}

\centering

\bibliographystyle{cas-model2-names}

\bibliography{cas-refs}

\end{document}